\documentclass[paper]{JHEP3}
\usepackage{epsfig}
\usepackage{amsmath}
\usepackage{amssymb}
\usepackage{amsbsy}
\usepackage{enumerate}
\usepackage{url}
\usepackage{xspace}

\newlength{\wfig}
\newlength{\hfig}
\newlength{\hfigs}
\newcommand{\tmop}[1]{\ensuremath{{\rm #1}}}

\newcommand{\HW}{{\tt HW}}
\newcommand{\HWJ}{{\tt HWJ}}
\newcommand{\HZ}{{\tt HZ}}
\newcommand{\HZJ}{{\tt HZJ}}

\newcommand{\HV}{{\tt HV}}
\newcommand{\HVJ}{{\tt HVJ}}

\newcommand\MH{M_{\scriptscriptstyle H}}

\newcommand\MV{M_{\scriptscriptstyle V}}
\newcommand\MHV{M_{\scriptscriptstyle V^*}}

\def\beq{\begin{equation}}
\def\beqn{\begin{eqnarray}}
\def\eeq{\end{equation}}
\def\eeqn{\end{eqnarray}}

\def\half{\frac{1}{2}}

\newcommand\NNLOPS{NNLO+PS}

\newcommand\HERWIG{{\tt HERWIG}}
\newcommand\HERWIGSIX{{\tt HERWIG}}
\newcommand\HWpp{{\tt HERWIG++}}
\newcommand\Herwigpp\HWpp
\newcommand\HERWIGPP\HWpp
\newcommand\PYTHIA{{\tt PYTHIA}}
\newcommand\PYTHIASIX{{\tt PYTHIA}\xspace}
\newcommand\PYTHIAEIGHT{{\tt Pythia8}\xspace}

\newcommand\MADGRAPH{{\tt MadGraph}}

\newcommand\KRA{K_{\scriptscriptstyle \rm R}}
\newcommand\KFA{K_{\scriptscriptstyle \rm F}}

\def\({\left(} 
\def\){\right)}

\newcommand\sss{\mathchoice%
{\displaystyle}%
{\scriptstyle}%
{\scriptscriptstyle}%
{\scriptscriptstyle}%
}

\newdimen\hbigcirc
\newdimen\wbigcirc

\newdimen\figwidth

\figwidth=\textwidth

\newcommand\as{\alpha_{\sss\rm S}}

\newcommand\pt{p_{\sss \rm T}}

\newcommand\kt{k_{\sss\rm T}}
\newcommand\qt{q_{\sss\rm T}}
\newcommand\qT{q_{\sss\rm T}}

\newcommand\CA{C_{\sss\rm A}}

\newcommand\CF{C_{\sss\rm F}}

\newcommand\POWHEG{{\tt POWHEG}}
\newcommand\POWHEGBOX{{\tt POWHEG BOX}}
\newcommand\GOSAM{{\tt GoSam}\xspace}
\newcommand\FORM{{\tt FORM}}
\newcommand\GOSAMCONT{{\tt GoSam-contrib}\xspace}
\newcommand\SAMURAI{{\tt SAMURAI}\xspace}
\newcommand\GOLEM{{\tt Golem95}\xspace}
\newcommand\HAGGIES{{\tt Haggies}}
\newcommand\MG{{\tt MadGraph4}}

\newcommand\MINLO{{\tt MiNLO}}
\newcommand\MiNLO{{\tt MiNLO}}

\newcommand\QGRAF{{\tt QGRAF}}

\newcommand\MGS{{\tt MadGraphStuff}}
\newcommand\gosamrc{{\tt gosam.rc}\xspace}
\newcommand\GSS{{\tt GoSamStuff}}
\newcommand\ttilde{\raise.17ex\hbox{$\scriptstyle\mathtt{\sim}$}}

\def\ord#1{{\cal O}\(#1\)}

\newcount\minutes 
\newcount\scratch 
\def\timestamp{% 
\scratch=\time 
\divide\scratch by 60 
\edef\hours{\the\scratch} 
\multiply\scratch by 60 
\minutes=\time 
\advance\minutes by -\scratch 
---$\,$\hours:\null 
\ifnum\minutes< 10 0\fi 
\the\minutes}

\preprint{MPP-2013-146, DSF-05-2013}

\title{$\boldsymbol{HW^\pm}$/$\boldsymbol{HZ}$ + 0 and 1 jet at NLO 
 with the POWHEG BOX interfaced to GoSam and their merging within MiNLO\vspace{-0.5cm}
}

\author{Gionata Luisoni\\ 
  Max-Planck Institut f{\"u}r Physik, F\"ohringer 6, D-80805 Munich, Germany\\
  E-mail: \email{luisonig@mpp.mpg.de}
}
\author{Paolo Nason\\
  INFN, Sezione di Milano Bicocca, Piazza della Scienza 3, 20126 Milano, Italy\\
  E-mail: \email{paolo.nason@mib.infn.it}
}

\author{Carlo Oleari\\
  Universit\`a di Milano-Bicocca and INFN, Sezione di Milano-Bicocca\\
  Piazza della Scienza 3, 20126 Milano, Italy\\
  E-mail: \email{carlo.oleari@mib.infn.it}}

\author{Francesco Tramontano\\
  Universit\`a di Napoli ``Federico II'' and INFN, Sezione di Napoli,\\
  Complesso di Monte Sant'Angelo, via Cintia, 80126 Napoli, Italy\\
  E-mail: \email{francesco.tramontano@na.infn.it}
}
\abstract{
  We present a generator for the production of a Higgs boson $H$ in
  association with a vector boson $V=W$ or $Z$ (including
  subsequent $V$ decay) plus zero and one jet, that can be used in
  conjunction with general-purpose shower Monte Carlo generators,
  according to the \POWHEG{} method, as implemented within the
  \POWHEGBOX{} framework.

  We have computed the virtual corrections using \GOSAM{}, a
  program for the automatic construction of virtual amplitudes. In
  order to do so, we have built a general interface of the
  \POWHEGBOX{} to the \GOSAM{} package. With this addition, the
  construction of a \POWHEG{} generator within the \POWHEGBOX{} is now
  fully automatized, except for the construction of the Born phase
  space.

  Our $HV +1$ jet generators can be run with the recently proposed \MINLO{}
  method for the choice of scales and the inclusion of Sudakov form factors.
  Since the $HV$ production is very similar to $V$ production, we were able
  to apply an improved \MINLO{} procedure, that was recently used in $H$ and
  $V$ production, also in the present case. This procedure is such that the
  resulting generator achieves NLO accuracy not only for inclusive
  distributions in $HV + 1$ jet production but also in $HV$ production,
  i.e.~when the associated jet is not resolved, yielding a further example of
  matched calculation with no matching scale.  }

\keywords{QCD, Hadronic Colliders, Higgs boson}
\begin{document}

\section{Introduction}
Higgs boson production in association with a vector boson ($HV$ production
from now on) is an interesting channel for Higgs boson studies at the LHC.
On one hand, it seems to be the only available channel to study the Higgs
branching to $b\bar{b}$, or to set limits to the Higgs branching into
invisible particles. In particular, in the $HV$ process with the Higgs boson
decaying into a $b\bar{b}$ pair, the CMS experiment has reported an excess of
events over the background of 2.2 standard deviation that is consistent with
an Higgs boson~\cite{CMS:cya}
%(CMS PAS HIG-12-044) 
%
in the 7~TeV data. The ATLAS experiment is not reporting any excess, but is
setting a limit above 1.9 standard deviation from the Standard Model
prediction~\cite{Aad:2012gxa}.
 %(ATLAS-CONF-2012-161) 
%
The CDF and D0 Collaborations have reported evidence for an excess of events,
at the 3.1 standard deviation level, in the search for the standard model
Higgs boson in the $HV$ process with the Higgs decaying to
$b\bar{b}$~\cite{Aaltonen:2012qt, Aaltonen:2012if, Abazov:2012qya}.  Looking
for invisible Higgs boson decays in $HZ$ associated production, the ATLAS
experiment is setting 95\%{} confidence level limits on a 125~GeV Higgs boson
decaying invisibly with a branching fraction larger than 65\%{}. Searches for
$WH\to WWW^{(*)}$ have also been carried out by both
ATLAS~\cite{ATLAS:2012toa}
%(ATLAS-CONF-2012-078) 
and CMS~\cite{CMS:zwa},
%(CMS PAS HIG-13-009)
and the Higgs boson decay into $\tau \bar{\tau}$ pairs has also been studied~\cite{CMS:xxa}.
%(CMS PAS HIG-12-051)

A \POWHEG{}~\cite{Nason:2004rx} generator for the $HV$ has been presented in
ref.~\cite{Hamilton:2009za}, and is often used for simulating the signal in
the experimental analysis.  It is developed within the \Herwigpp{}
framework~\cite{Bahr:2008pv}.

The aim of this paper is
twofold:
\begin{enumerate}
\item to present generators for the $HV$ and $HV$ + 1 jet processes in the
  \POWHEGBOX{}~\cite{Frixione:2007vw, Alioli:2010xd} framework, a
  next-to-leading order+parton shower~(NLO+PS) event generators.  In the
  following we will refer to these generators as \HV{} and \HVJ{}, respectively.
  These generators can be interfaced with any parton shower compliant
  with the Les Houches Interface for User
  Processes~\cite{Boos:2001cv,Alwall:2006yp},
  like \PYTHIA{}~\cite{Sjostrand:2006za}, 
  \PYTHIAEIGHT{}~\cite{Sjostrand:2007gs},
  \HERWIG~\cite{Corcella:2002jc} and \Herwigpp{}~\cite{Bahr:2008pv}.
\item To illustrate a new interface of the \POWHEGBOX{} to the
  \GOSAM{}~\cite{Cullen:2011ac} package, that allows for the automatic
  generation of the virtual amplitudes.  In order to achieve this, the
  \POWHEGBOX{} interface to \MG{} of ref.~\cite{Campbell:2012am} was extended
  to produce also a file that can be passed to \GOSAM{} in order to generate
  the virtual amplitudes.  Using this new tool, the generation of all matrix
  elements is performed automatically, and one only needs to supply the Born
  phase space in order to build a \POWHEG{} process.
\end{enumerate}
In our $HV$ + 1 jet generators,
we apply the improved version of the \MINLO{} procedure~\cite{Hamilton:2012np}
discussed in ref.~\cite{Hamilton:2012rf}.
In~\cite{Hamilton:2012rf} it was shown that, by applying
this procedure to the NLO production of a color-neutral object in association
with one jet, one can reach NLO accuracy for quantities that are inclusive in
the production of the color-neutral system, i.e.~when the associated jet is
not resolved.  In the present case, the $HVj$ process can be viewed as the
production of a virtual vector boson, that decays into the $HV$ pair,
accompanied by one jet. Since vector-boson production was explicitly
considered in ref~\cite{Hamilton:2012rf}, the same \MINLO{} procedure used in
that context can be transported and applied to the present case. The
\HVJ{}+\MINLO{} generator that we build can then replace the \HV{} generator,
since it has the same NLO accuracy, and in addition it is NLO accurate in the
production of the hardest jet. We are then able to produce a matched
calculation, with no matching scale, without actually merging different
samples.

In addition to this, it turns out that it is possible to extend the precision
of our \HVJ{}+\MINLO{} generator in such a way to reach
next-to-next-to-leading order+parton shower (NNLO+PS) accuracy for inclusive
$HV$ distributions.  This can be achieved following the procedure outlined in
ref.~\cite{Hamilton:2012rf}, i.e.~by rescaling the \HVJ{}+\MINLO{} results
with the the NNLO calculation for $HV$ production, already available in the
literature~\cite{Ferrera:2011bk}.  We remind that higher accuracy for the
production of an associated jet is particularly useful in contexts where an
extra jet is vetoed, like in the search for invisible Higgs boson decays, or,
in general, when events are also classified according to the number of jets.
In this paper, we will not pursue the extension of our calculation to the
NNLO+PS accuracy, postponing a phenomenological study of this to a future
publication.

The organization of the paper is the following: in sec.~\ref{sec:gosam} we
describe the new \GOSAM{} - \POWHEGBOX{} interface. In sec.~\ref{sec:virtual}
we give more details about the virtual contribution and in
sec.~\ref{sec:minlo} we briefly review the \MINLO{} procedure for the present
case. In sec.~\ref{sec:pheno} we compare the improved \HVJ{}+\MINLO{} outputs
with the \HV{} ones, and discuss a few phenomenological results. Finally in
sec.~\ref{sec:conclusions} we summarize our findings. Instructions on how to
generate a new virtual code using the \GOSAM{} - \POWHEGBOX{} interface and
how to drive the \GOSAM{} program are collected in
Appendixes~\ref{app:gosam_how_to} and~\ref{app:input_card}.

\section{The \GOSAM{} - \POWHEGBOX{} interface}
\label{sec:gosam}
The code for the calculation of the one-loop corrections to $HVj$ production
has been generated using \GOSAM{} interfaced to the
\POWHEGBOX{}. \GOSAM{} is a
python framework coupled to a template system for the automatic generation of
{\tt fortran95} codes for the evaluation of virtual amplitudes. The one-loop
virtual corrections are evaluated using algebraic expressions of
$D$-dimensional amplitudes based on Feynman diagrams. The diagrams are
initially generated using \QGRAF{}~\cite{Nogueira:1991ex}. \GOSAM{} allows to
select the relevant diagrams which are processed with
\FORM~\cite{Kuipers:2012rf}, using the \texttt{SPINNEY}~\cite{Cullen:2010jv}
package. The processing amounts to organize the numerical computation of the
amplitudes in terms of their numerators that are functions of the loop
momentum.  Finally, the manipulated algebraic expressions of the one-loop
numerators are optimized and converted into a {\tt fortran95}
code\footnote{This phase in \GOSAM{} can be performed by \FORM{} (version 4.0
  or higher) or through the JAVA program \HAGGIES~\cite{Reiter:2009ts}.} and
merged into the code generated by \GOSAM{}.

When integrating over the phase space, the virtual matrix elements are
evaluated with \SAMURAI~\cite{Mastrolia:2010nb} or, alternatively, with
\GOLEM~\cite{Cullen:2011kv}. The first program evaluates the amplitude using
integrand reduction methods~\cite{Ossola:2006us} extended to
$D$-dimensions~\cite{Ellis:2007br}, whereas the latter allows to compute the
same amplitude evaluating tensor-integrals. The interplay of the two
reduction strategies is used to guarantee the highest speed for the produced
codes, while keeping the required precision for good numerical stability of
the results. For the evaluation of the scalar one-loop integrals {\tt
  QCDLoop}~\cite{vanOldenborgh:1990yc,Ellis:2007qk}, {\tt
  OneLOop}~\cite{vanHameren:2010cp} or {\tt Golem95C}~\cite{Cullen:2011kv}
can be used.

\subsection{Binoth-Les-Houches-Accord interface}
\GOSAM{} has an interface to generic external Monte Carlo~(MC) programs based
on the Binoth-Les-Houches-Accord~(BLHA)~\cite{Binoth:2010xt}, which sets the
standards for the communications among a MC program and a general
One-Loop Program~(OLP). We developed an interface based on the BLHA for the
\POWHEGBOX{} MC program as well, and the computation we present here is its
first application. In the following we explain the basic features of this
interface.

The communication between MC and OLP in the BLHA has two separate stages: a
pre-running phase and a running-time phase~\cite{Binoth:2010xt}.

\begin{enumerate}
\item
In the pre-running phase, during code generation, the MC writes an
\textit{order file} which contains all the basic information about the
amplitudes that should be generated and computed by the OLP. Among these
there are the powers of the strong and electromagnetic couplings, the type of
corrections (i.e.~whether the OLP should generate QCD, QED or EW loop
corrections), information on the helicity and color treatment (average,
sum\ldots), and finally the full list of the partonic subprocesses for which
the virtual one-loop amplitudes are required.  The OLP reads the order file
and generates the needed amplitudes together with the code to evaluate
them. Furthermore, information on the generated code are written by the OLP
into a \textit{contract file}, which will be read-in again by the MC at every
run. The structure of the contract file is similar to the one of the order
file. In the former the requests of the MC, contained in the latter, are
either confirmed by an ``OK'' label, or rejected. This allows to be sure that
the requests contained in the order file can be satisfied and the codes are
coherent, or to detect potential problems due to misunderstanding of the MC
requests by the OLP. On top of this, the list of partonic processes is
rewritten in the contract file with a numerical label which uniquely
identifies each subprocess. This will be used at running time by the MC to
obtain the amplitude of a specific subprocess from the OLP.

\item
At running time the MC reads-in once the contract file and then communicate
with the OLP via two standard subroutines: the first one is responsible for
the initialization of the OLP, whereas the second one is called for the
evaluation of the virtual corrections of a specific partonic subprocess at a
given phase-space point. As a reply, the OLP provides an array containing the
coefficients of the poles and the finite part of the virtual contribution.
\end{enumerate}
We have extended the already existing interface~\cite{Campbell:2012am} of the
\POWHEGBOX{} to \MG{} in order to write an order file and read a contract
file.

Together with the order file, \GOSAM{} needs a further input card: this is a
file, called \gosamrc, where the user can specify further details on the
generation of the virtual contributions. Among these there are, for example,
the number of available cpus or the treatment of classes of gauge-invariant
diagrams.

This setup allows for a completely automated generation of all the matrix
elements needed by the \POWHEGBOX{} for the computation of the QCD
corrections to any Standard Model process. The limitation is, of course, the
computing power of nowadays computer.

In Appendix~\ref{app:gosam_how_to} the reader can find a detailed
descriptions of the steps needed for the generation of a new process, whereas
a list of options for the \GOSAM{} input card can be found in
Appendix~\ref{app:input_card}.

\section{The virtual corrections}
\label{sec:virtual}
In this section we would like to comment on the structure of the virtual
diagrams contributing to $HVj$ production.
\begin{figure}[htb]
\begin{center}
\includegraphics[width=0.48\textwidth]{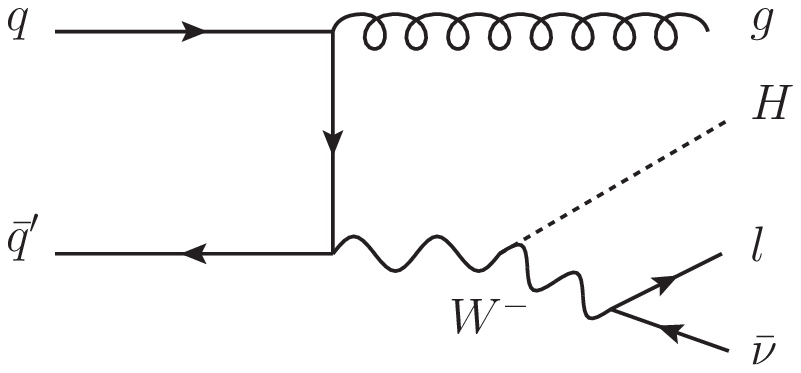}
\includegraphics[width=0.48\textwidth]{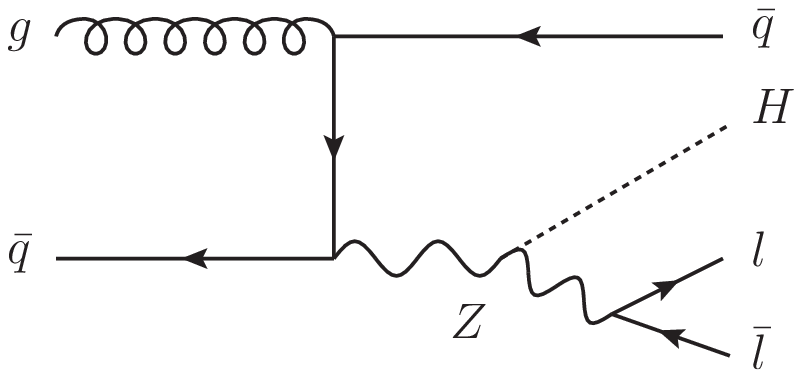}
\end{center}
\caption{A sample of leading-order Feynman diagrams for $HWj$ and $HZj$
  production.}
\label{fig:tree} 
\end{figure}
Typical tree-level diagrams for the $HWj$ and $HZj$ production are
illustrated in fig.~\ref{fig:tree}. Similar ones can be drawn for
real-radiation diagrams. In all these contributions, the Higgs boson is
radiated off the vector boson. In fact, we consider all quark to be massless,
with the only exception for the top quark, running in fermionic loops.

The virtual corrections can be separated into three different classes:
\begin{enumerate}[(a)]

\item \label{item:loop} In the first class, we can accommodate the one-loop
  Higgs-Strahlung-type diagrams, with no closed fermionic loop.  
  \begin{figure}[htb]
    \begin{center}
      \includegraphics[width=0.48\textwidth]{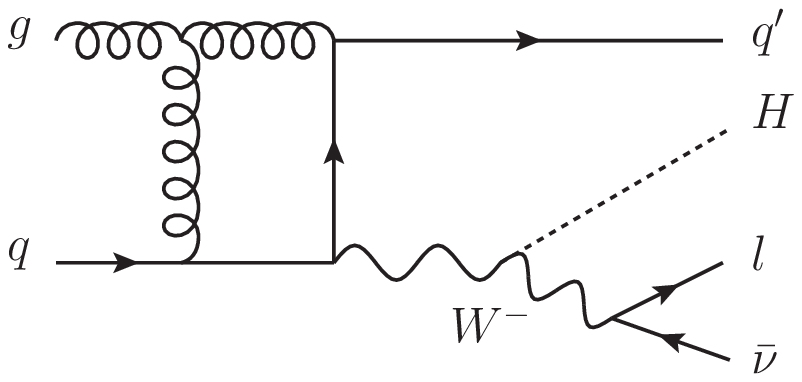}
      \includegraphics[width=0.48\textwidth]{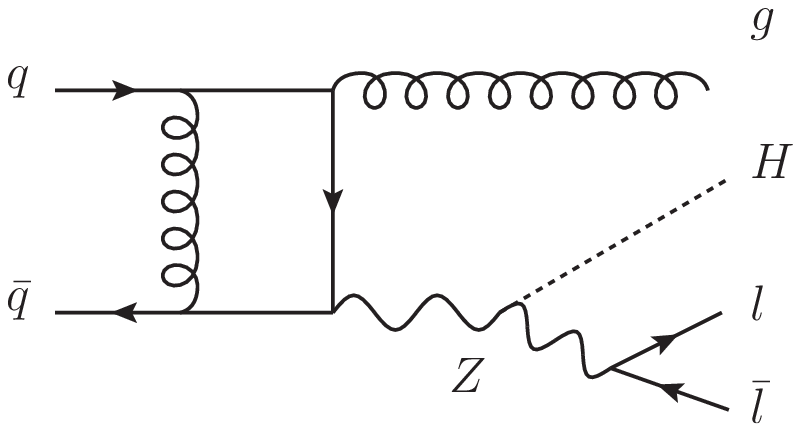}
    \end{center}
    \caption{A sample of one-loop Higgs-Strahlung diagrams, with no closed
      fermionic loop.}
    \label{fig:loop} 
  \end{figure}
  A sample of diagrams belonging to this class is depicted in
  fig.~\ref{fig:loop}.  These diagrams are similar to the virtual diagrams
  for $H/W/Z$ production, with the addition of an extra parton in the final
  state.

\item \label{item:furry} A sample of diagrams belonging to the second class
  of virtual corrections are illustrated in fig.~\ref{fig:furry}.
  \begin{figure}[htb]
    \begin{center}
      \includegraphics[width=0.48\textwidth]{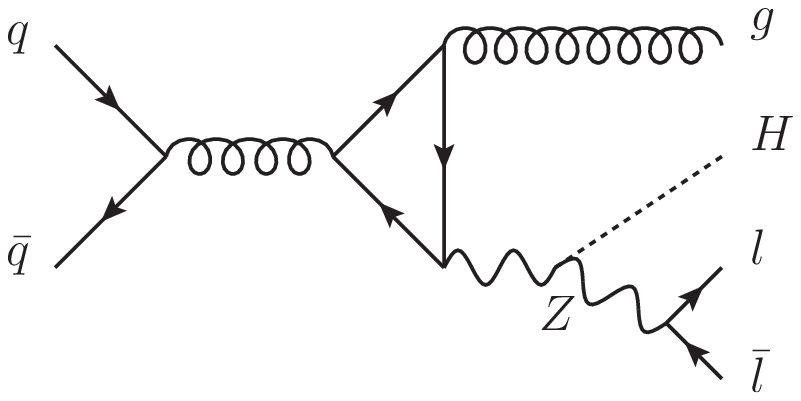}
      \includegraphics[width=0.48\textwidth]{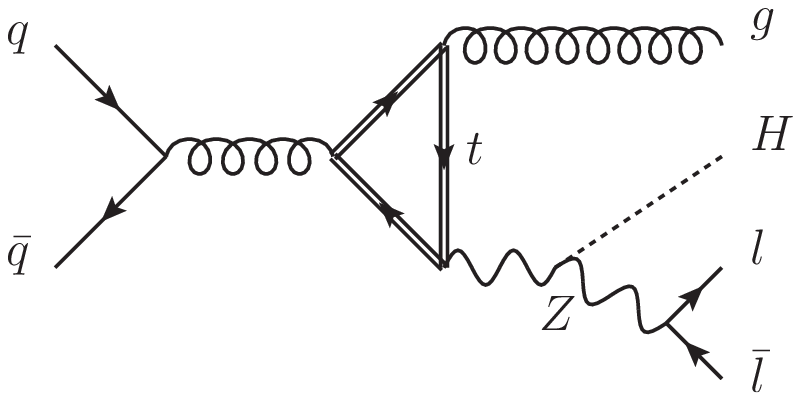}
    \end{center}
    \caption{A sample of one-loop Higgs-Strahlung diagrams, with massless and
      massive closed fermionic loop.}
    \label{fig:furry} 
  \end{figure}
  In this figure we have plotted Higgs-Strahlung-type diagrams when a closed
  quark loop is present, and the $Z$ boson couples to the internal quark. No
  such diagrams are present for $W$ production, since the flavour running in
  the loop must be conserved.  These contributions vanish by
  charge-conjugation invariance (Furry's theorem), when they couple to a
  vector current.
  For axial currents, they cancel in pairs of up-type and down-type quarks,
  because they have opposite axial coupling, as long as the loop of different
  flavours can be considered massless.  Thus, the up-quark contribution
  cancels with the down-quark, and, since we treat the charm as massless, its
  contribution cancels with the strange one. Only the difference between the
  diagrams with a massive top quark and a massless bottom quark loop
  survives.

\item \label{item:Yt} In the last class, we have the Feynman diagrams where
  the Higgs boson couples directly to the massive top-quark loop. A sample of
  this type of diagrams is illustrated in figs.~\ref{fig:topdiags}
  and~\ref{fig:topdiags1}.
  \begin{figure}[htb]
    \begin{center}
      \includegraphics[width=0.48\textwidth]{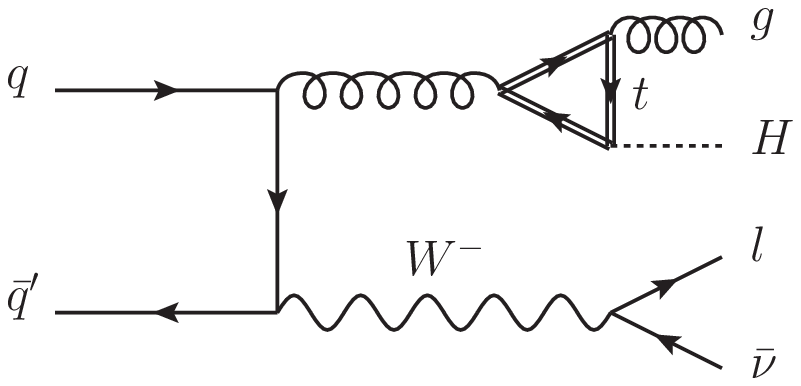}
      \includegraphics[width=0.48\textwidth]{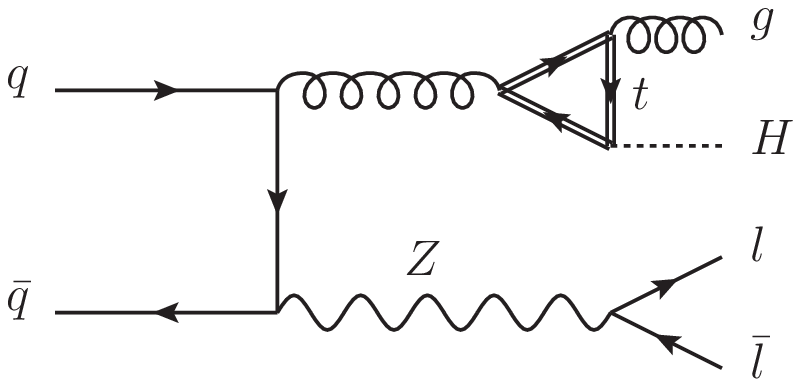}
    \end{center}
    \caption{A sample of virtual diagrams involving a massive top-quark loop,
      where the Higgs boson couples directly to the top quark.}
    \label{fig:topdiags} 
  \end{figure}

  \begin{figure}[htb]
    \begin{center}
      \includegraphics[width=0.50\textwidth]{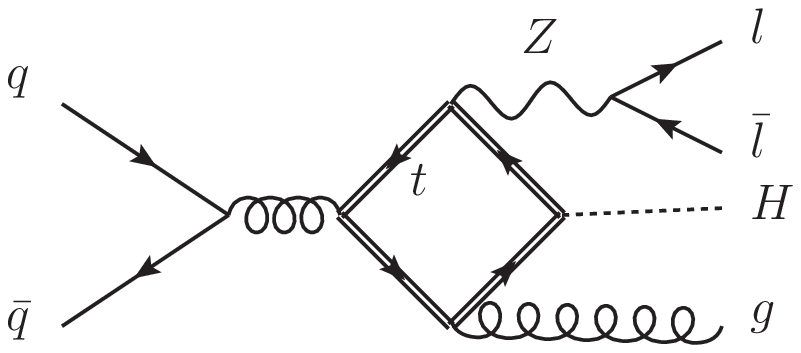}
    \end{center}
    \caption{A sample of virtual diagrams involving a massive top-quark loop,
      where the Higgs boson couples directly to the top quark.}
    \label{fig:topdiags1} 
  \end{figure}

\end{enumerate}
The Feynman diagrams belonging to the three classes are fully implemented in
the \POWHEGBOX{} code, and they are computed if the {\tt massivetop} flag is
set to 1 in the input file.

%HZJ NLO 20 GeV, antikt 0.5  
%totsig
%massless = $0.518744E-002 \pm  0.361453E-005$  pb 
%massive = $0.525429E-002 \pm  0.362188E-005$

The contributions of the diagrams belonging to classes~(\ref{item:furry})
and~(\ref{item:Yt}) are, in general, very small.  For example, in $HZj$
production, with the setup described in sec.~\ref{sec:pheno} and with
transverse-momentum cuts on jets of 20~GeV, the total NLO cross section,
keeping only the virtual diagrams belonging to class~(\ref{item:loop}), is
5.187(4)~fb, while keeping all the virtual diagrams is 5.254(4)~fb.  This
behavior is reflected in more exclusive quantities, such as the
transverse momentum and rapidity of the $HZ$ pair or of the $H$ and $Z$
bosons.
\begin{figure}[htb]
\begin{center}
\includegraphics[width=0.49\textwidth]{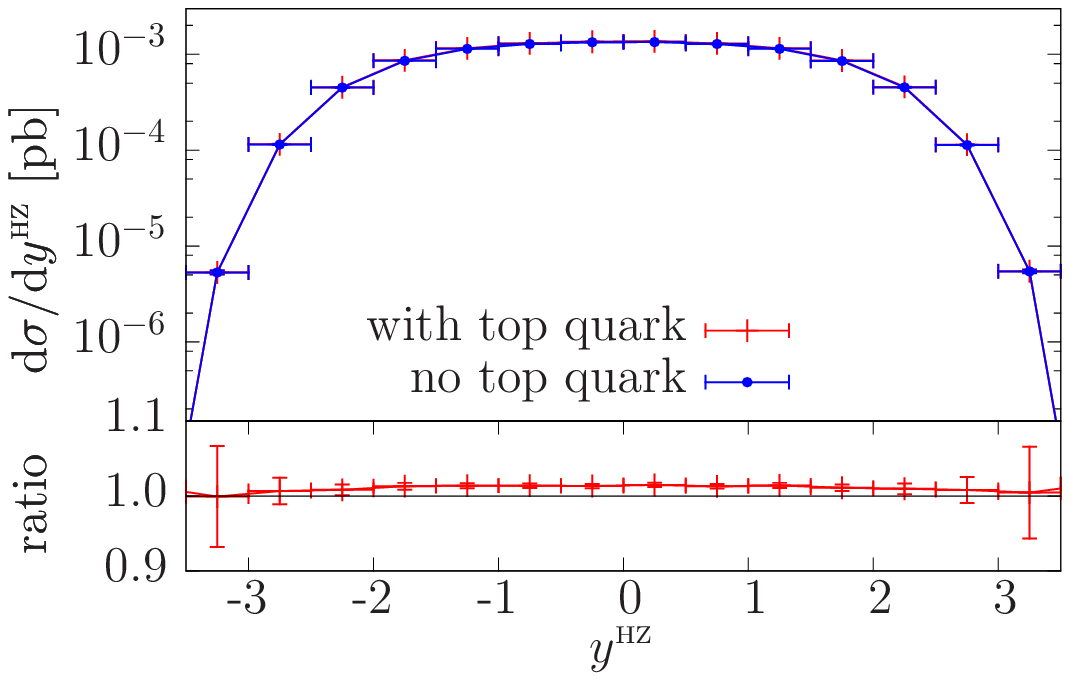}
\includegraphics[width=0.49\textwidth]{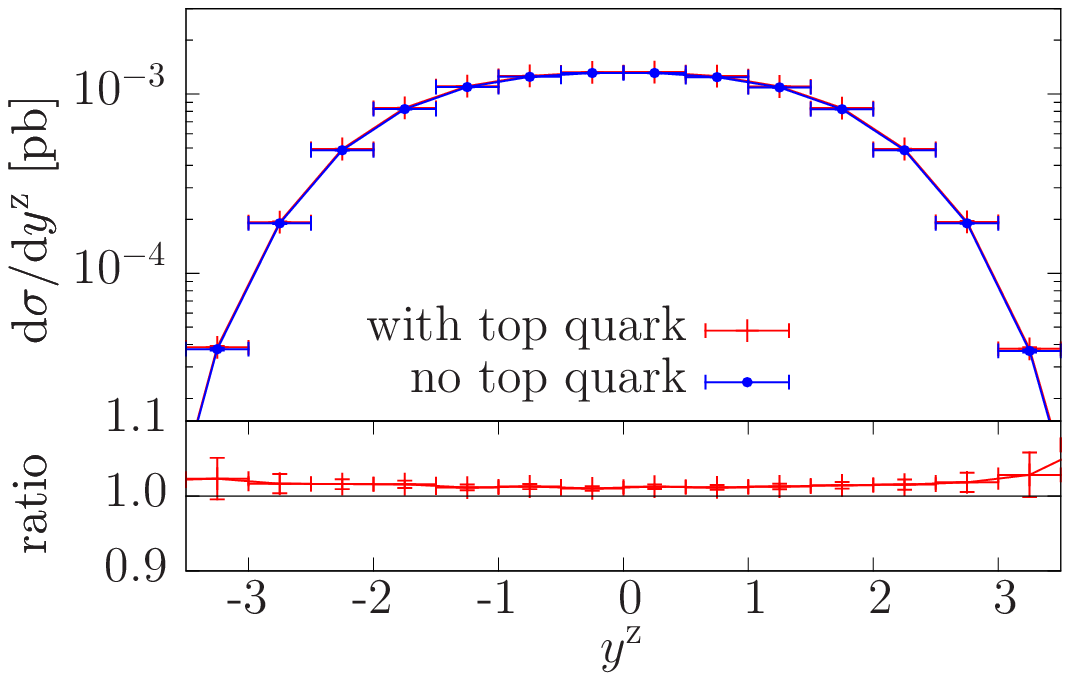}
\end{center}
\caption{NLO rapidity distributions of the $HZ$ pair (left plot) and of the
  $Z$ boson (right plot), in $HZj$ production. The red curves were obtained
  by using the full set of virtual diagrams, including the Feynman graphs
  containing a top-quark loop. The blue curves were computed neglecting the
  diagrams belonging to classes~(\ref{item:furry}) and~(\ref{item:Yt}).}
\label{fig:NLO_mass_y} 
\end{figure}
\begin{figure}[htb]
\begin{center}
\includegraphics[width=0.49\textwidth]{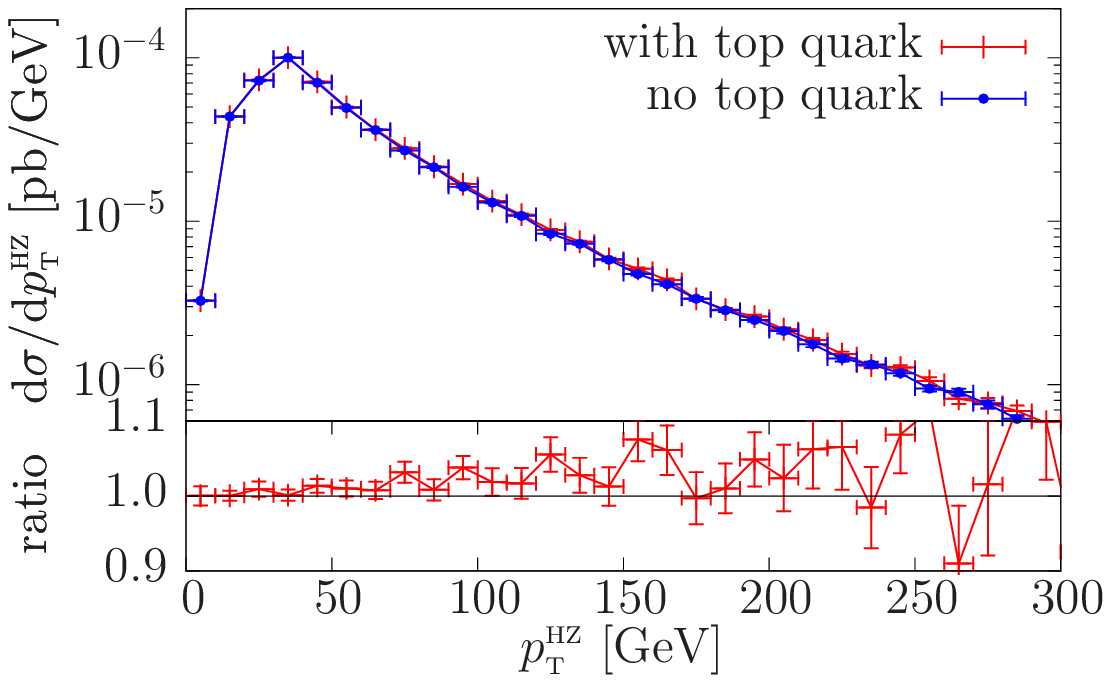}
\includegraphics[width=0.49\textwidth]{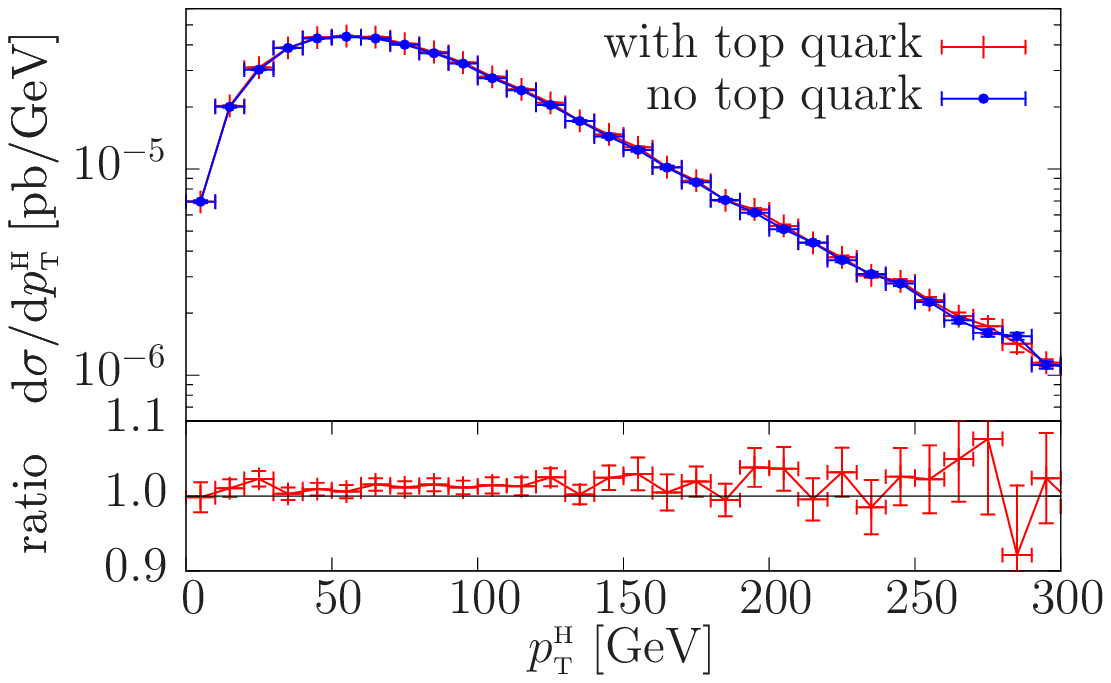}
\end{center}
\caption{NLO transverse-momentum distributions of the $HZ$ pair (left plot)
and of the $H$ boson (right plot), in $HZj$ production. The labels are as in
fig.~\protect{\ref{fig:NLO_mass_y}}.}
\label{fig:NLO_mass_pt} 
\end{figure}
In fig.~\ref{fig:NLO_mass_y}, we compare the rapidity distributions of the
$HZ$ system (left plot) and of the $Z$ boson (right plot), obtained by
including the virtual diagrams with the top-quark loop and by neglecting
them.  In fig.~\ref{fig:NLO_mass_pt} we show a similar comparison for the
transverse-momentum distributions of the $HZ$ pair and of the Higgs boson.

In all the several observables that we have examined, we find differences of
the order of 1-2\%, with the exception of distributions related to the $HZ$
transverse momentum (or of the leading jet $\pt$), that display a slightly
larger difference increasing with the transverse momentum. Observe that, in
this case, our generator would still include correctly the effect of the
diagrams of the classes~(\ref{item:furry}) and~(\ref{item:Yt}). In fact, the
\MINLO{} correction affects their contribution only at small transverse
momentum, where they are negligible (see sec.~\ref{sec:minlo} for more details).

Since the contributions of the diagrams belonging to
classes~(\ref{item:furry}) and~(\ref{item:Yt}) are at the level of a few
percent for inclusive and typical more exclusive distributions, the default
behavior of the \POWHEGBOX{} is to neglect them, i.e.~the default value of
the {\tt massivetop} flag is 0.
We can then apply in a straightforward way the improved \MINLO{} procedure to
$HVj$ production, as illustrated in sec.~\ref{sec:minlo}.  We would like to
point out that the diagrams belonging to classes~(\ref{item:furry})
and~(\ref{item:Yt}) not only have a small impact on the cross sections, but
they contribute to the differential cross section with terms that are finite,
down to zero transverse momentum of the jet, i.e.~they do not have the
diverging behavior of the diagrams belonging to class~(\ref{item:loop}).

\section{The \MINLO{} procedure}
\label{sec:minlo}
The application of the \MINLO{} procedure to $HVj$ production is fully
analogous to the case of the $Vj$ generator presented in
ref.~\cite{Hamilton:2012rf}, and based on ref.~\cite{Hamilton:2012np}, if we
keep only the virtual diagrams that belong to class~(\ref{item:loop}),
i.e.~if the production mechanism is an Higgs-Strahlung one
\begin{equation}
p p \,\to\, V^* j\,, \qquad {\rm with} \qquad V^* \,\to\, H V \,\to\, H\,
l_1\, l_2\,,
\end{equation}
with no top-quark loop involved. 

In the \MINLO{} method, the NLO inclusive cross section for the computation
of the underlying Born kinematics (the so called $\bar{B}$ function in the
\POWHEG{} jargon) is modified with the inclusion of the Sudakov form factor
and with the use of appropriate scales for the couplings, according to the
formula
\begin{equation}
\label{eq:sudagen}
\bar{B}= \as \left( \qT \right) \Delta^2 \!\left( \MHV ,
  \qT \right) \left[ B \left( 1 - 2 \Delta^{\left( 1 \right)} \!\left(
    \MHV, \qT \right) \right) + V + \int d \Phi_{\tmop{rad}} \,
    R\right], 
\end{equation}
where $\MHV$ is the virtuality of the vector boson before the Higgs-boson
emission, i.e.~$\MHV^2=(p_{\sss l_1}+p_{\sss l_2}+p_{\rm \sss H})^2$, where
$p_{\sss l_1}$ and $p_{\sss l_2}$ are the momenta of the leptons into which
the $V$ boson decays, and $p_{\rm \sss H}$ is the Higgs boson momentum.  The
transverse momentum of $V^*$ is indicated with $\qt$.  In
eq.~(\ref{eq:sudagen}) we have stripped away one power of $\as$ from the Born
($B$), the virtual ($V$) and the real ($R$) contribution, and we have
explicitly written it in front, with its scale dependence. The scale at
which the remaining power of $\as$ in $R$, $V$ and $\Delta^{\left( 1
  \right)}$ is evaluated and the factorization scale used in the evaluation
of the parton distribution functions is again  $\qt$.
The Sudakov form factor $\Delta$ is given by
\begin{equation}
\label{eq:gluon_sudakov}
  \Delta \left( Q, \qT \right) = \exp \left\{ - \int_{\qT^2}^{Q^2}
  \frac{d q^2}{q^2} \left[ A \left( \as \left( q^2 \right) \right) \log
    \frac{Q^2}{q^2} + B \left( \as \left( q^2 \right) \right) \right]
  \right\},
\end{equation}
and
\begin{equation}
\label{eq:deltag1}
  \Delta \left( Q, \qT \right) = 1 + \Delta^{\left( 1 \right)}
  \left( Q, \qT \right) + \mathcal{O} \left( \as^2 \right)
\end{equation}
is the expansion of $\Delta$ in powers of $\as$. 
The functions $A$ and $B$ have a perturbative expansion in terms of constant
coefficients
\begin{equation}
\label{eq:A_and_B}
A \left( \as \right) = \sum_{i = 1}^{\infty} A_i \, \as^i\,, \hspace{2em} B \left(
\as \right) = \sum_{i = 1}^{\infty} B_i \,\as^i \,.
\end{equation}
In the improved \MiNLO{} approach, only the coefficients $A_1$, $A_2$, $B_1$
and $B_2$ are needed in order to have NLO accuracy also in inclusive $HV$
distributions.  Their value for the case at hand are given
by~\cite{Kodaira:1981nh,Davies:1984hs,Davies:1984sp}
\begin{eqnarray}
\label{eq:AiBiq}
A_1 &=& \frac{1}{2\pi}\CF, \quad  
A_2 =\frac{1}{4\pi^2}\CF\,K, \quad 
B_1 = -\frac{3}{4\pi}\CF\,,\quad 
K = \left(\frac{67}{18}-\frac{\pi^2}{6}\right)\CA-\frac{5}{9}n_f\,,
\\
\label{eq:B2q}
B_2&=& \frac{1}{2\pi^2} \left[
\( \frac{\pi^2}{4}-\frac{3}{16}-3\zeta_3\) \CF^2 + \(\frac{11}{36} \pi^2
-\frac{193}{48} +\frac{3}{2}\zeta_3\)\CF\CA  +
\(\frac{17}{24}-\frac{\pi^2}{18}\)\CF n_f \right]
\nonumber \\
&&\mbox{}+ 4\zeta_3 (A_1)^2\,,
\end{eqnarray}
and the $\ord{\as}$ expansion of the Sudakov form factor in
eq.~(\ref{eq:deltag1}) is given by
\begin{equation}
 \Delta^{\left( 1 \right)}\!\left( Q, \qT \right)= \as \left[ 
-\frac{1}{2} A_1 \log^2\frac{\qt^2}{Q^2} + B_1 \log\frac{\qt^2}{Q^2} \right].
\end{equation}
Following the reasoning in ref.~\cite{Hamilton:2012rf} we can show that
events generated according to eq.~(\ref{eq:sudagen}), i.e.~our \HVJ-\MINLO{}
generator, are NLO-accurate for distributions inclusive in the $HV$
production and have NLO accuracy for distributions inclusive in the $HV$ + 1
jet too.

\section{Implementation and plots }
\label{sec:pheno}
In this section we discuss and compare results obtained using the \HV{} and
the \HVJ{}-\MiNLO{} generator, as implemented in the \POWHEGBOX{}.

In our study we have generated 5 millions events both for the $HVj$ and for
the $HV$ sample, where $V$ is a $W^-$, a $W^+$ or a $Z$ boson, which decays
leptonically.  The conclusions drawn for associated $W^+$ production are
similar to those for $W^-$ production. For this reason, in the following, we
will show only results for $W^-$ production.

The produced samples were generated for the LHC running at 8~TeV, with
$\MH=125$~GeV and $\Gamma_H= 4.03$~MeV, and with the Higgs boson virtuality
distributed according to a fixed-width Breit-Wigner function.  In addition,
we have restricted the Higgs boson and $V$ boson virtuality in the range
10~GeV--1~TeV. This range can be set by the user via the {\tt powheg.input}
file.  A minimum transverse momentum cut of 260~MeV has been applied to the
jet in the $HVj$ sample in the generation of the underlying Born kinematics,
in order to avoid the Landau pole in the strong coupling constant.  The
factorization scale for the \HV{} \POWHEG{} generators has been set to $\MH +
\MV$. The renormalization and factorization scales for the \HVJ+\MINLO{}
generators have been set according to the procedure discussed in
sec.~\ref{sec:minlo}.
In our study, we have used the CT10 parton distribution function
set~\cite{Lai:2010vv}, but any other set can be used
equivalently~\cite{Martin:2009iq, Ball:2012cx}.  The shower has been
completed using the \PYTHIASIX{} shower Monte Carlo program, although it is
as easy to interface our results to \HERWIGSIX, \HERWIGPP{} and
\PYTHIAEIGHT{}. Jets have been reconstructed using the anti-$\kt$ jet
algorithms, as implemented in the {\tt fastjet}
package~\cite{Cacciari:2005hq, Cacciari:2008gp} with $R=0.5$.

The \HV{} code is run without using the {\tt hfact} flag, that can be used to
separate the real contribution into a singular and a finite part.  In the
present case, since radiative corrections are modest, we do not expect a
large sensitivity to this parameter, as is observed in Higgs boson production
in gluon fusion~\cite{Alioli:2008tz, Dittmaier:2012vm}. In the following, the
\HVJ{} generator is always run in the \MINLO{} mode.

\begin{table}[htb]
\centering
{\footnotesize
\begin{tabular}{|c|c|c|c|c|c|c|c|c|c|}
\hline
\multicolumn{8}{|c|}{$HW^-\to H l^- \bar{\nu}_l$ production total cross sections in fb at the LHC, 8~TeV}\\ \hline
 $K_R,K_F$  & $1,1$ & $1,2$ & $2,1$ & $1,\half$ & $\half,1$ & $\half,\half$ & $2,2$ \\ \hline
\HWJ{}-\MiNLO{}  
& $ 28.10(4)$ & $ 28.44(4)$ & ${\bf  27.27(3)}$ & $ 27.68(4)$ & $ 28.53(8)$ & $
{\bf 29.05(8)}$ & $ 27.75(3)
$\\ \hline
\HW{}
& $ 28.5578(7)$ & $ 28.7418(1)$ & ${\bf 28.1927(7)}$ & $ 28.4269(9)$ & ${\bf 28.9993(9)}$ & $ 28.881(1)$ & $ 28.3691(9)
$\\ \hline
 \end{tabular}
}
\caption{Total cross section for $HW^-\to H l^- \bar{\nu}_l$ at the 8~TeV
  LHC, obtained with the \HWJ{}-\MiNLO{} and the \HW{} programs, at NLO
  level, for different scales combinations.  The maximum and minimum of the
  cross sections are highlighted.}
\label{tab:totHW} 
\end{table}

\begin{table}[htb]
\centering
{\footnotesize
\begin{tabular}{|c|c|c|c|c|c|c|c|c|c|}
\hline
\multicolumn{8}{|c|}{$HZ\to H e^+ e^-$ production total cross sections in fb at the LHC, 8~TeV}\\ \hline
 $K_R,K_F$  & $1,1$ & $1,2$ & $2,1$ & $1,\half$ & $\half,1$ & $\half,\half$ & $2,2$ \\ \hline
\HZJ{}-\MiNLO{}  
& $ 12.818(9)$ & $ {\bf 12.478(7)}$ & $ 12.97(1)$ & $ 12.93(2)$ & $
12.659(9)$ & $ {\bf 13.14(2)}$ & $ 12.684(9)
$\\ \hline
\HZ{}
& $ 13.0979(4)$ & $ {\bf 12.9304(4)}$ & $ 13.1705(5)$ & $ 13.3002(5)$ & $
13.0501(4)$ & $ {\bf 13.2559(4)}$ & $ 12.9986(4)
$\\ \hline
 \end{tabular}
}
\caption{Total cross section for $HZ\to H e^+ e^-$ at the 8~TeV LHC, obtained
  with the \HZJ{}-\MiNLO{} and the \HZ{} programs, at NLO level, for
  different scales combinations.  The maximum and minimum of the
  cross sections are highlighted.}
\label{tab:totHZ} 
\end{table}

\begin{figure}[htb]
\begin{center}
\includegraphics[width=0.49\textwidth]{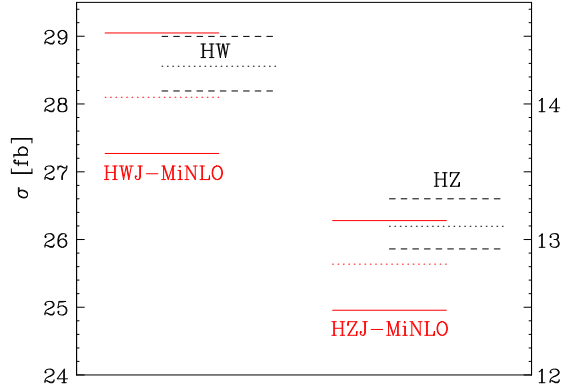}
\end{center}
\caption{Total cross section variation for \HVJ{}-\MiNLO{} (solid red) and
  \HV{} (dashed black). The maximum and minimum values for the total cross
  section are taken from tabs.~\ref{tab:totHW} and~\ref{tab:totHZ}. The total
  cross section with central scales is drawn in dotted lines.}
\label{fig:HVJ_sigtot} 
\end{figure}

Since the improved \MINLO{} prescription applied here achieves NLO accuracy
for observables inclusive in the $HV$ production, we begin showing results
for the most inclusive quantity, i.e.~the total cross section.  In
tabs.~\ref{tab:totHW} and~\ref{tab:totHZ} we collect the results for the
total cross sections obtained with the \HVJ{}-\MiNLO{} and the \HV{}
programs, both at full NLO level, for different scale combinations.  The
scale variation in the \HVJ{}-\MiNLO{} results is obtained by multiplying the
factorization scale and each of the several renormalization scales that
appear in the procedure by the scale factors $\KFA$ and $\KRA$, respectively,
where
\begin{equation}
(\KRA,\KFA)=(0.5,0.5),  (0.5,1), (1,0.5), (1,1),(2,1),(1,2),(2,2).
\end{equation}
The Sudakov form factor is also changed according to the prescription
described in ref.~\cite{Hamilton:2012rf}.  For ease of visualization, in
fig.~\ref{fig:HVJ_sigtot} we have plotted the maximum and minimum values for
the \HVJ-\MiNLO{} (red lines) and \HV{} (black lines) cross sections in solid
and dashed lines, respectively. We have also plotted the central-scale cross
section in dotted lines.
Notice that we expect agreement only up to terms of higher order in $\as$,
since the \HVJ-\MiNLO{} results include terms of higher order, and also since
the meaning of the scale choice is different in the two approaches. For
similar reasons, we do not expect the scale variation bands to be exactly the
same in the two approaches.  From the tables and the figure, it is clear that
the standard \HV{} NLO+PS results and the \HVJ{}-\MiNLO{} one are fairly
consistent: the \HVJ{}-\MiNLO{} independent scale variation is in general
larger than the \HV{} one, and it shrinks if a symmetric scale variations is
performed, as illustrated in the last two columns of the tables.  In general
the \HVJ{}-\MiNLO{} central values are 2\% smaller than the \HV{} ones.
As already pointed out in ref.~\cite{Hamilton:2012rf}, comparing full
independent scale variation in the \HVJ-\MiNLO{} and in the \HV{} approaches
does not seem to be totally fair. In fact, in the \HV{} case, there is no
renormalization scale dependence at LO, while there is such a dependence in
\HVJ-\MiNLO{}. It was shown in ref.~\cite{Hamilton:2012rf} for the case of
$W$ production at LO that an independent scale variation corresponds at least
in part to a symmetric scale variation in the \MiNLO{} formula. It is thus
not surprising that the \MiNLO{} independent scale variation is so much
larger than the \HV{} one also at NLO. If we limit ourselves to consider
only symmetric scale variations, the \MiNLO{} and the \HV{} results are more
consistent, although the \HV{} scale variation band is extremely small.

\begin{figure}[htb]
\begin{center}
\includegraphics[width=0.49\textwidth]{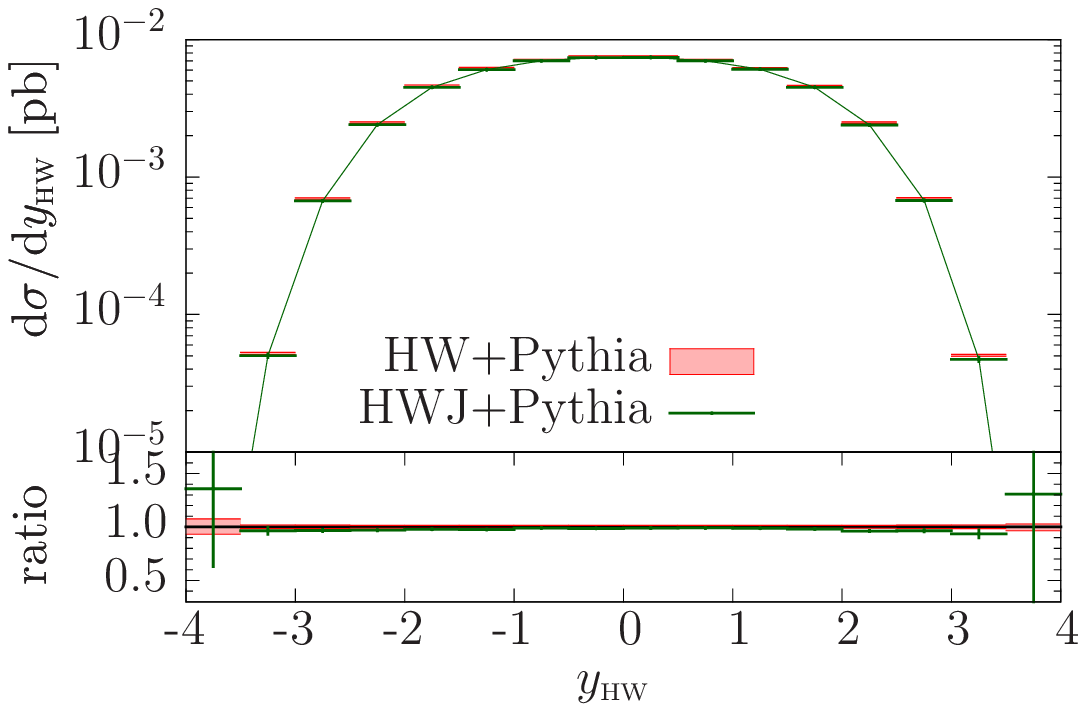}
\includegraphics[width=0.49\textwidth]{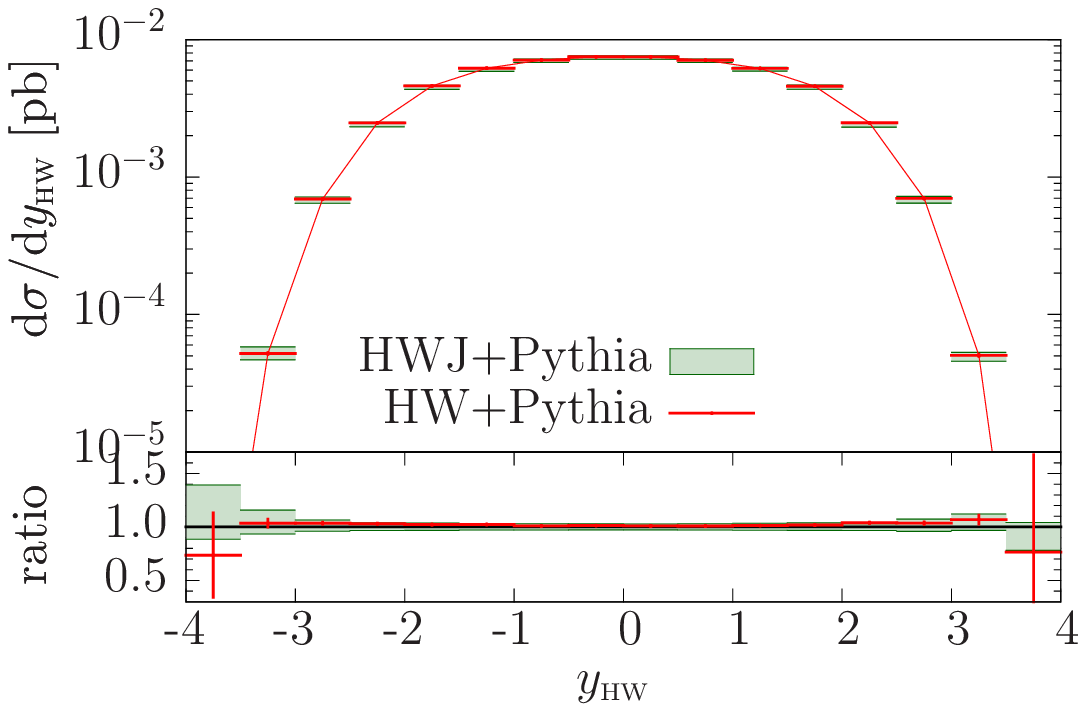}
\end{center}
\caption{Comparison between the \HW+\PYTHIA{} result and the
  \HWJ-\MiNLO+\PYTHIA{} result for the $HW^-$ rapidity distribution at
  the LHC at 8~TeV.  The left plot shows the 7-point scale-variation band for
  the \HW{} generator, while the right plot shows the \HWJ-\MiNLO{} 7-point
  band.}
\label{fig:HWy} 
\end{figure}
Turning now to less inclusive quantities, we plot in fig.~\ref{fig:HWy} the
rapidity distribution of the $HW$ system obtained with the \HW{} and
\HWJ{}-\MiNLO{} generator. We remind that this quantity is predicted at NLO
by both generators, and in fact the agreement is very good. The uncertainty
band of the \HW{} generator is shown on the left while that of the
\HWJ{}-\MiNLO{} generator is shown on the right.

\begin{figure}[htb]
\begin{center}
\includegraphics[width=0.49\textwidth]{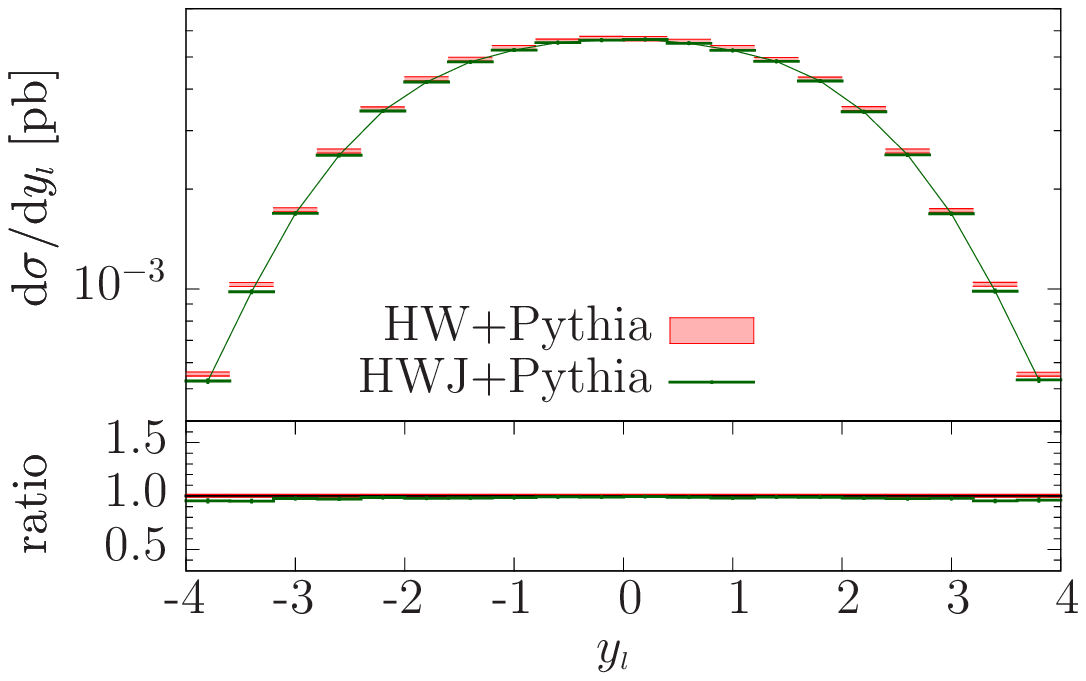}
\includegraphics[width=0.49\textwidth]{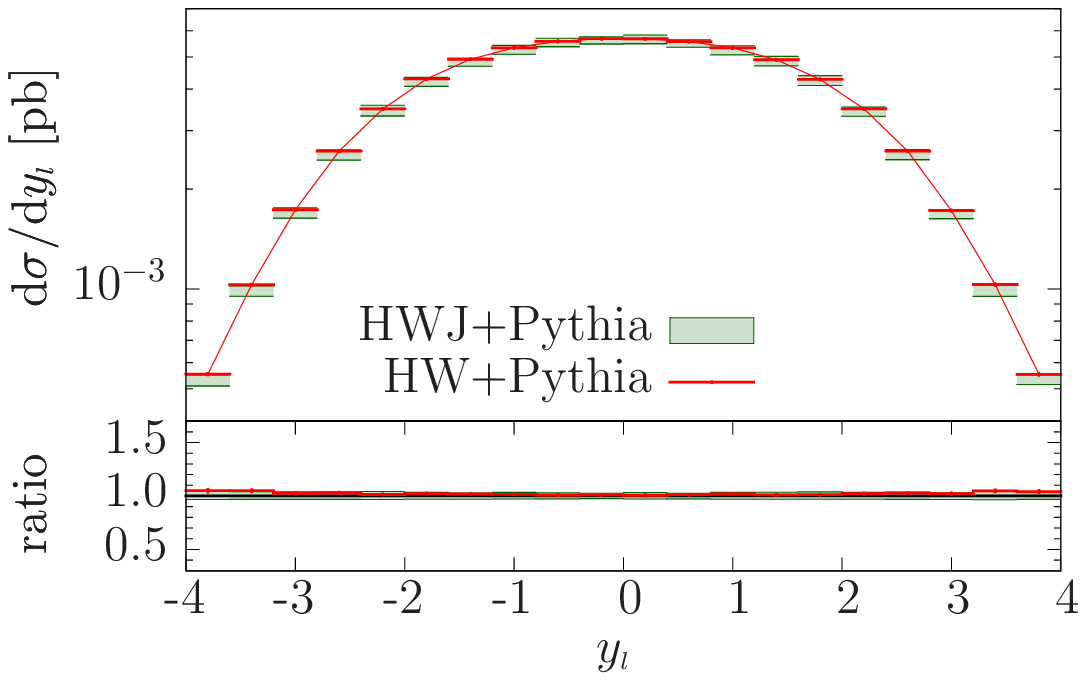}
\end{center}
\caption{Comparison between the \HW+\PYTHIA{} result and the
  \HWJ-\MiNLO+\PYTHIA{} result for the rapidity distribution of the charged
  lepton from the $W^-$ decay, at the LHC at 8~TeV.  The left plot shows the
  7-point scale-variation band for the \HW{} generator, while the right plot
  shows the \HWJ-\MiNLO{} 7-point band.}
\label{fig:HW_ly} 
\end{figure}
In fig.~\ref{fig:HW_ly} we show another inclusive quantity, i.e.~the charged
lepton transverse momentum from the $W^-$ decay. Also in this case we find
perfect agreement between the two generators

\begin{figure}[htb]
\begin{center}
\includegraphics[width=0.49\textwidth]{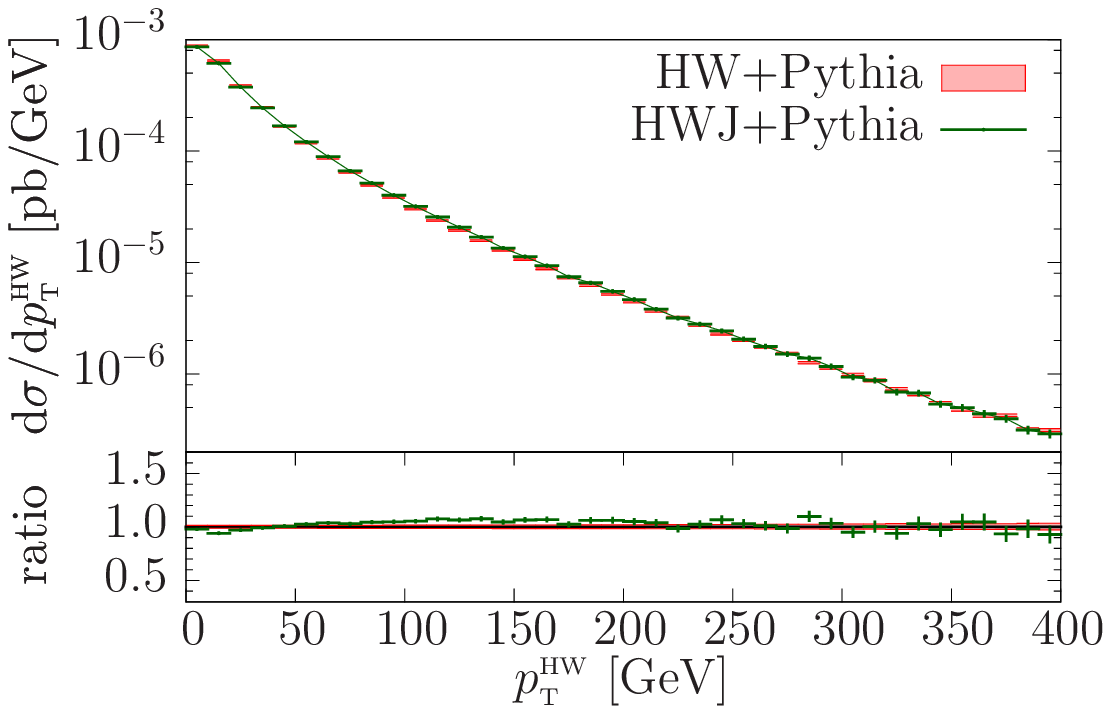}
\includegraphics[width=0.49\textwidth]{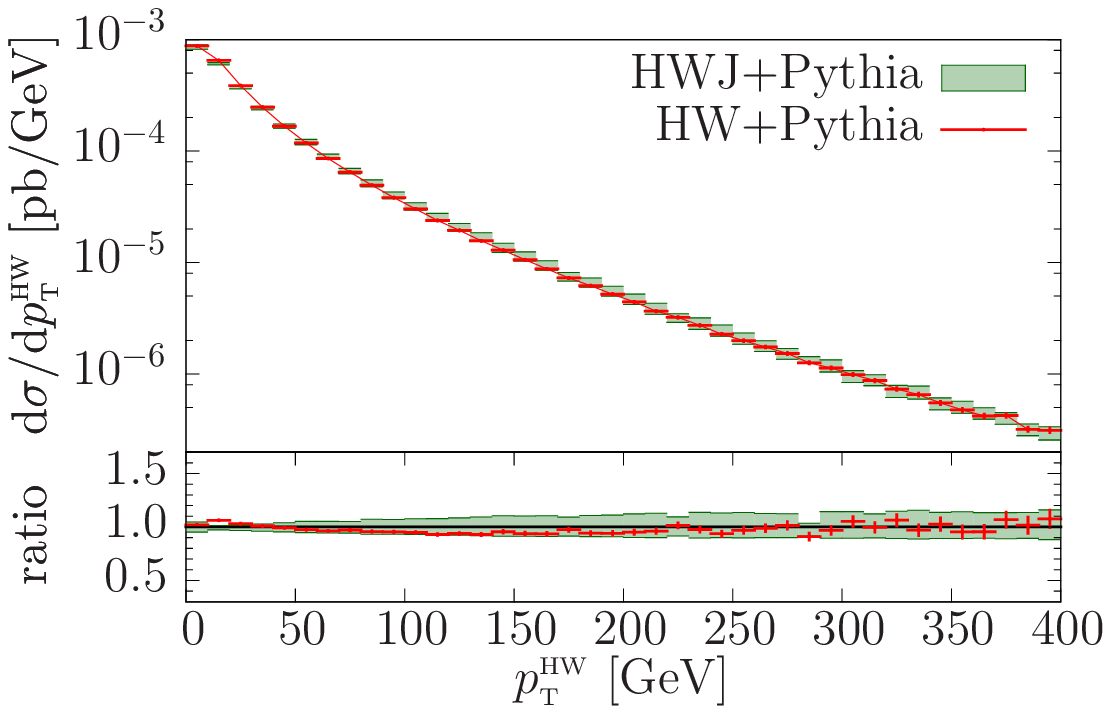}
\end{center}
\caption{Comparison between the \HW+\PYTHIA{} result and the
  \HWJ{}-\MiNLO{}+\PYTHIA{} result for the $HW^-$ transverse-momentum
  distribution. The bands are obtained as in fig.~\ref{fig:HWy}.}
\label{fig:HWpt} 
\end{figure}

\begin{figure}[htb]
\begin{center}
\includegraphics[width=0.49\textwidth]{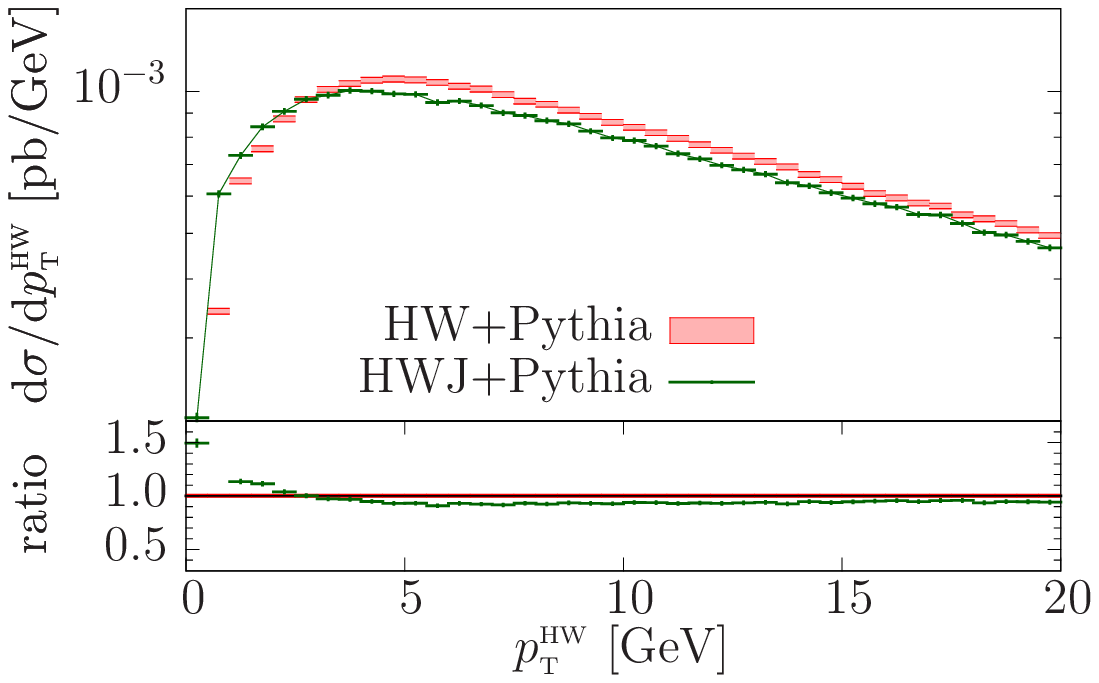}
\includegraphics[width=0.49\textwidth]{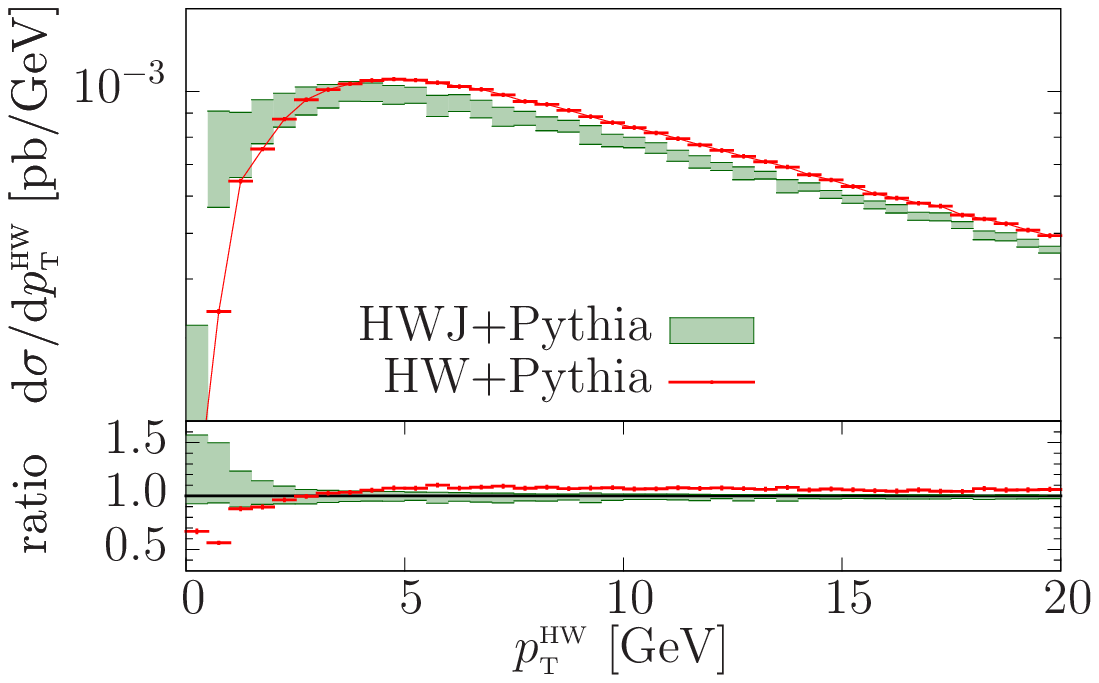}
\end{center}
\caption{Same as fig.~\ref{fig:HWpt} for a different $\pt^{\rm\sss HW}$ range.}
\label{fig:HWptzm} 
\end{figure}
In figs.~\ref{fig:HWpt} and~\ref{fig:HWptzm} we compare the \HW{} and
\HWJ{}-\MiNLO{} generators for the transverse momentum of the $HW$ system.
In this case we do observe small differences, that are however perfectly
acceptable if we remember that this distribution is only computed at leading
order by the \HW{} generator, while it is computed at NLO accuracy by the
\HWJ{}-\MiNLO{} generator. It can also be noted that the uncertainty band for
the \HW{} generator is uniform, while it depends upon the transverse momentum
for the \HWJ{}-\MiNLO{} one.  In fact, the uniformity of the scale-variation
band in the \HW{} case is well understood: in \POWHEG{}, the scale
uncertainty manifests itself only in the $\bar{B}$ function, while the shape
of the transverse-momentum distribution is totally insensitive to it.

\begin{figure}[htb]
\begin{center}
\includegraphics[width=0.49\textwidth]{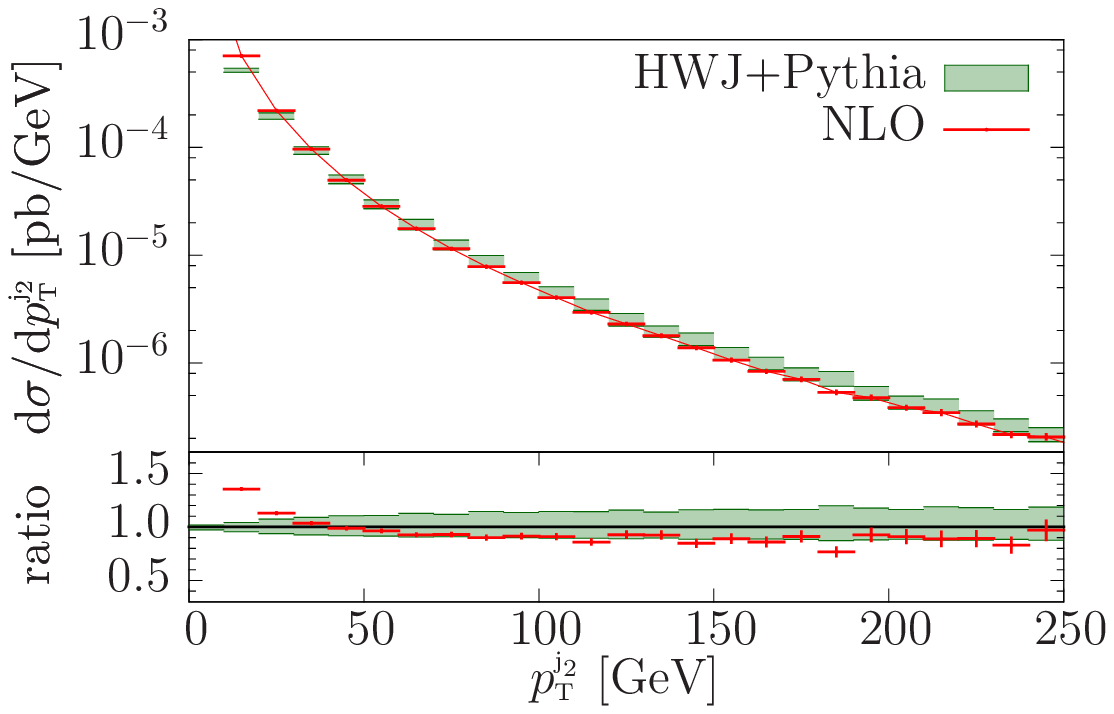}
\includegraphics[width=0.49\textwidth]{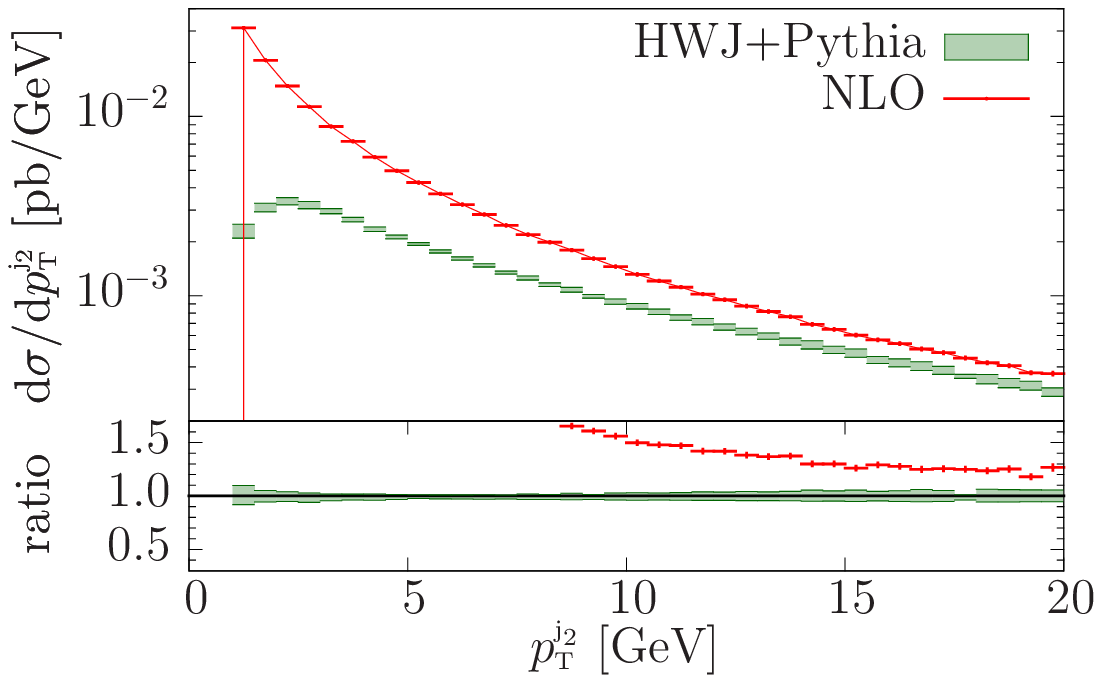}
\end{center}
\caption{Comparison between the \HWJ-\MiNLO+\PYTHIA{} and the NLO \HWJ{} result
  for the transverse momentum of the second hardest jet, at the LHC at 8~TeV,
  in two different $\pt$ ranges.  The plots shows the 7-point scale-variation
  band for the \HWJ{} generator.}
\label{fig:HW_j2pt} 
\end{figure}
The transverse momentum of the second jet computed with the \HWJ{}-\MiNLO{}
generator compared with the pure NLO result is plotted in
fig.~\ref{fig:HW_j2pt}. In this plot, \MINLO{} plays no role, but the
\POWHEG{} formalism is still in place. In fact, the NLO prediction for the
second jet has a diverging behavior at low transverse momenta, that is tamed
in the \POWHEGBOX{} generator by the Sudakov form factor.  Thus, the two
results differ considerably from each other, especially at low transverse
momenta.

\begin{figure}[htb]
\begin{center}
\includegraphics[width=0.49\textwidth]{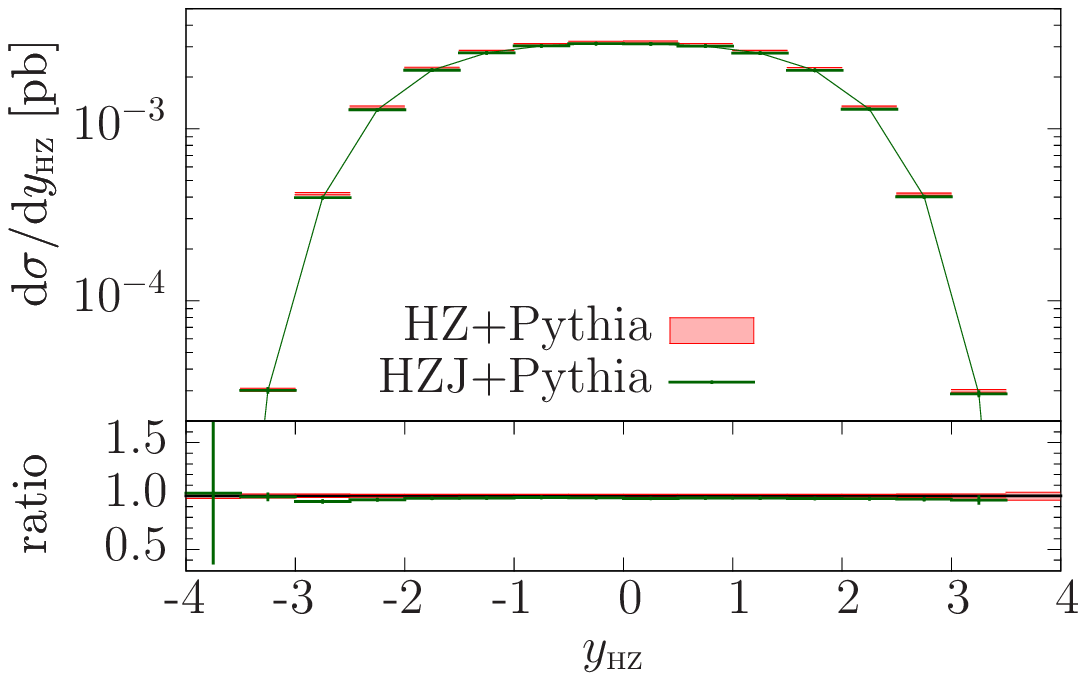}
\includegraphics[width=0.49\textwidth]{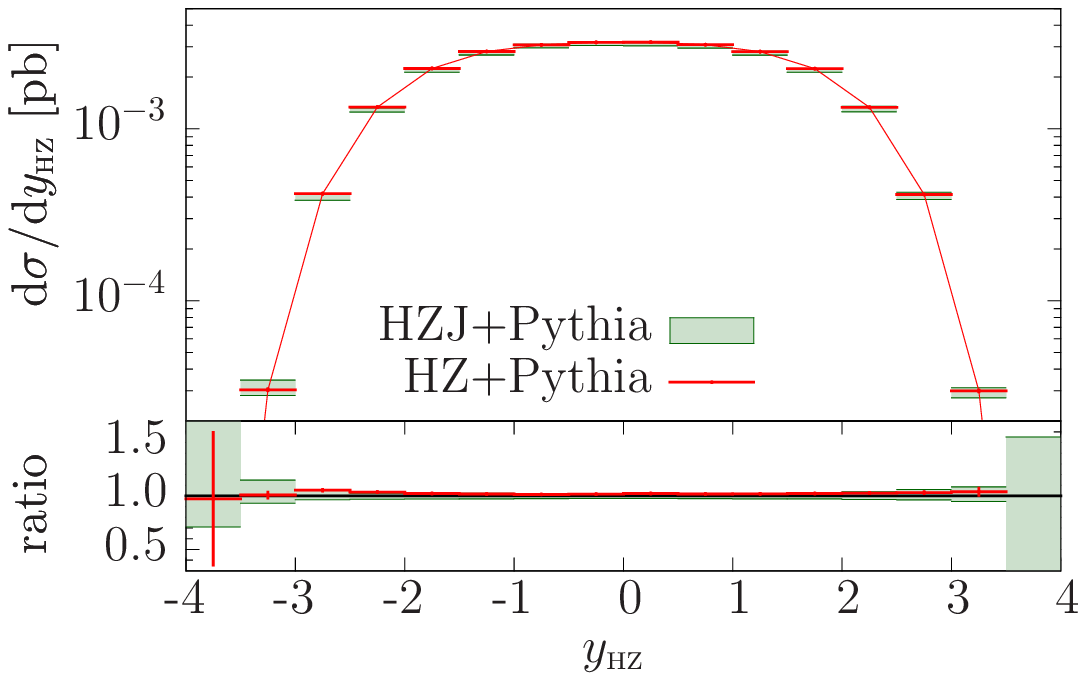}
\end{center}
\caption{Comparison between the \HZ+\PYTHIA{} result and the
  \HZJ-\MiNLO+\PYTHIA{} result for the $HZ$ rapidity distribution at
  the LHC at 8~TeV.  The left plot shows the 7-point scale-variation band for
  the \HZ{} generator, while the right plot shows the \HZJ-\MiNLO{} 7-point
  band.}
\label{fig:HZy} 
\end{figure}

\begin{figure}[htb]
\begin{center}
\includegraphics[width=0.49\textwidth]{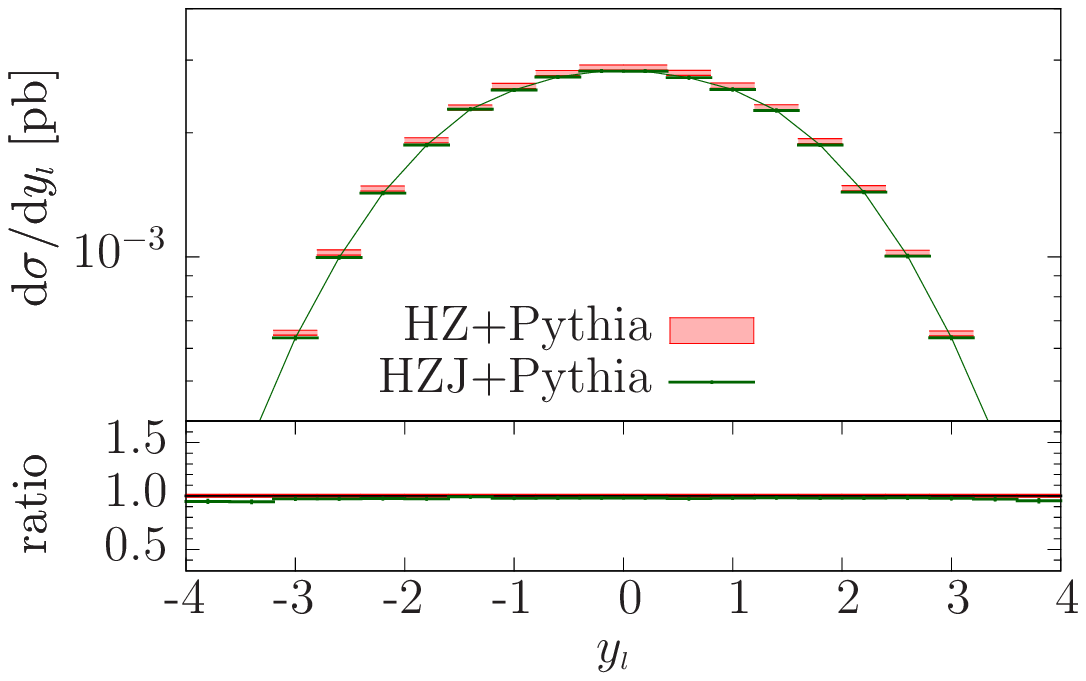}
\includegraphics[width=0.49\textwidth]{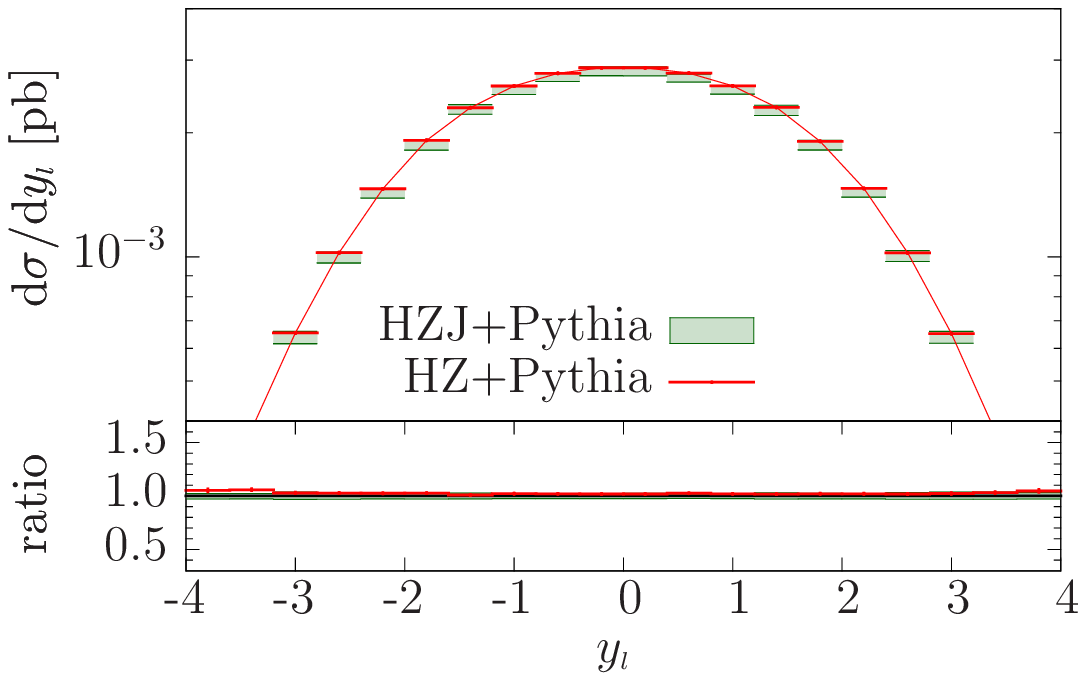}
\end{center}
\caption{Comparison between the \HZ+\PYTHIA{} result and the
  \HZJ-\MiNLO+\PYTHIA{} result for the rapidity distribution of the electron
  from the $Z$ decay, at the LHC at 8~TeV.  The left plot shows the 7-point
  scale-variation band for the \HZ{} generator, while the right plot shows
  the \HZJ-\MiNLO{} 7-point band.}
\label{fig:_HZ_ly} 
\end{figure}

\begin{figure}[htb]
\begin{center}
\includegraphics[width=0.49\textwidth]{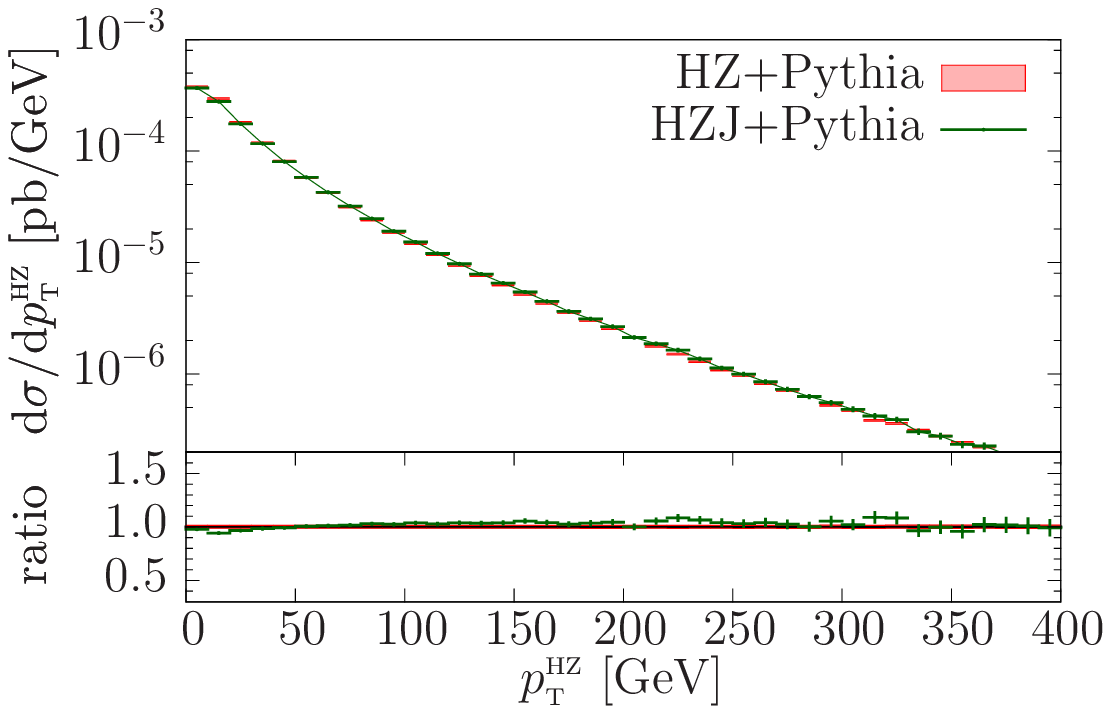}
\includegraphics[width=0.49\textwidth]{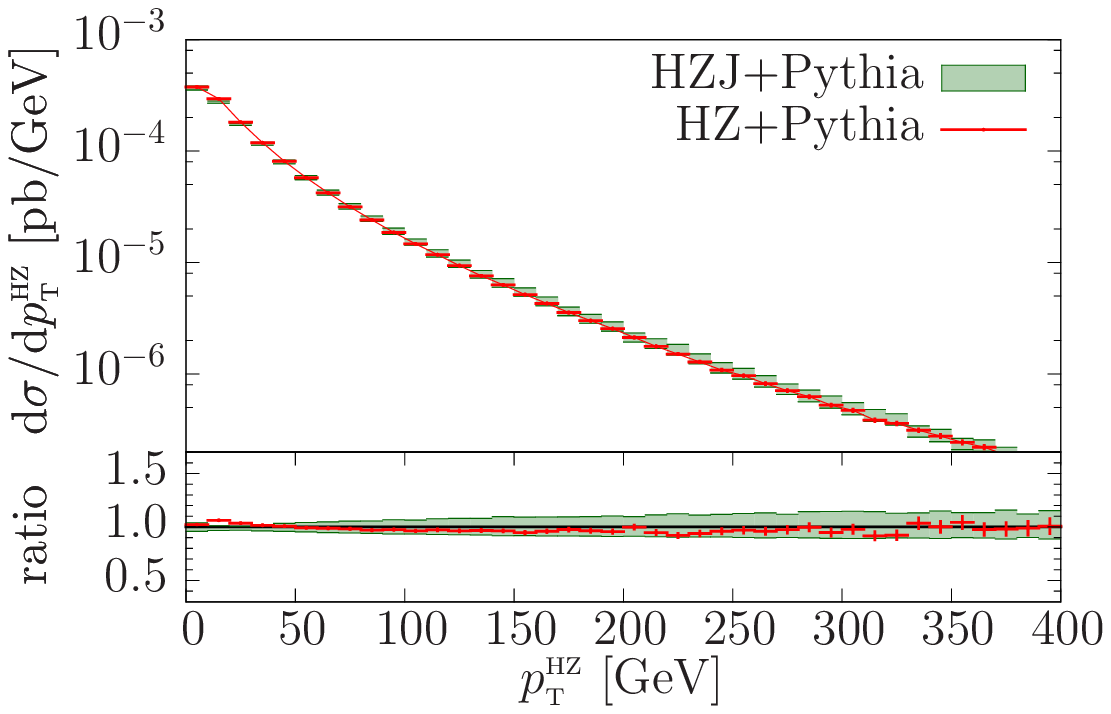}
\end{center}
\caption{Comparison between the \HZ+\PYTHIA{} result and the
  \HZJ{}-\MiNLO{}+\PYTHIA{} result for the $HZ$ transverse-momentum
  distribution. The bands are obtained as in fig.~\ref{fig:HZy}.}
\label{fig:HZpt} 
\end{figure}

\begin{figure}[htb]
\begin{center}
\includegraphics[width=0.49\textwidth]{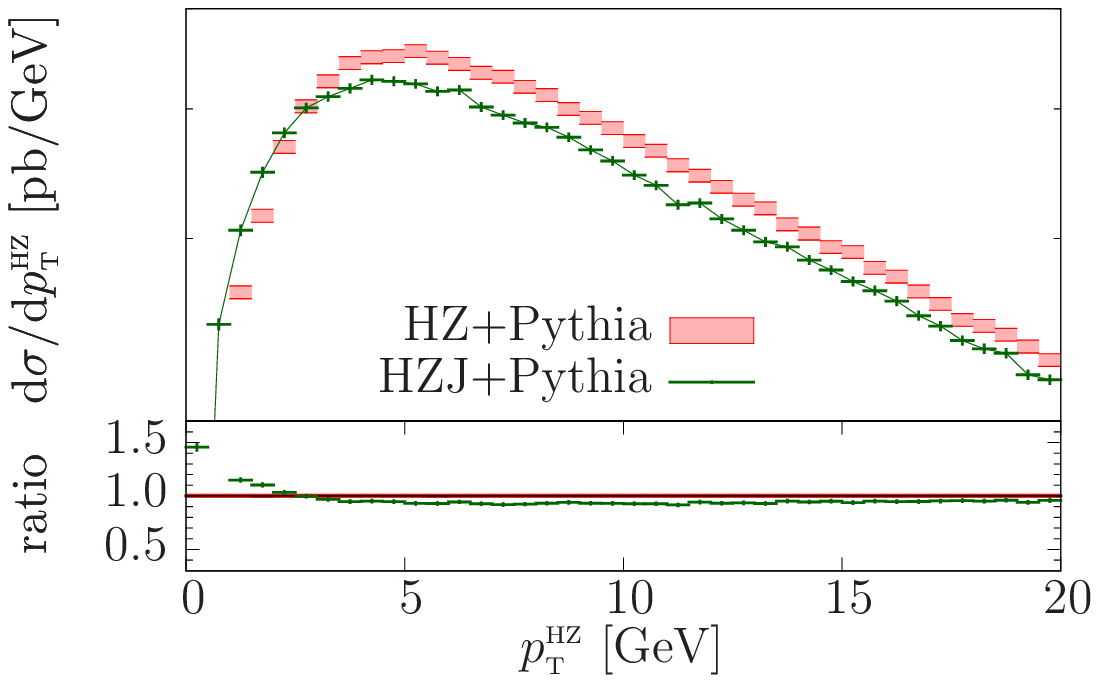}
\includegraphics[width=0.49\textwidth]{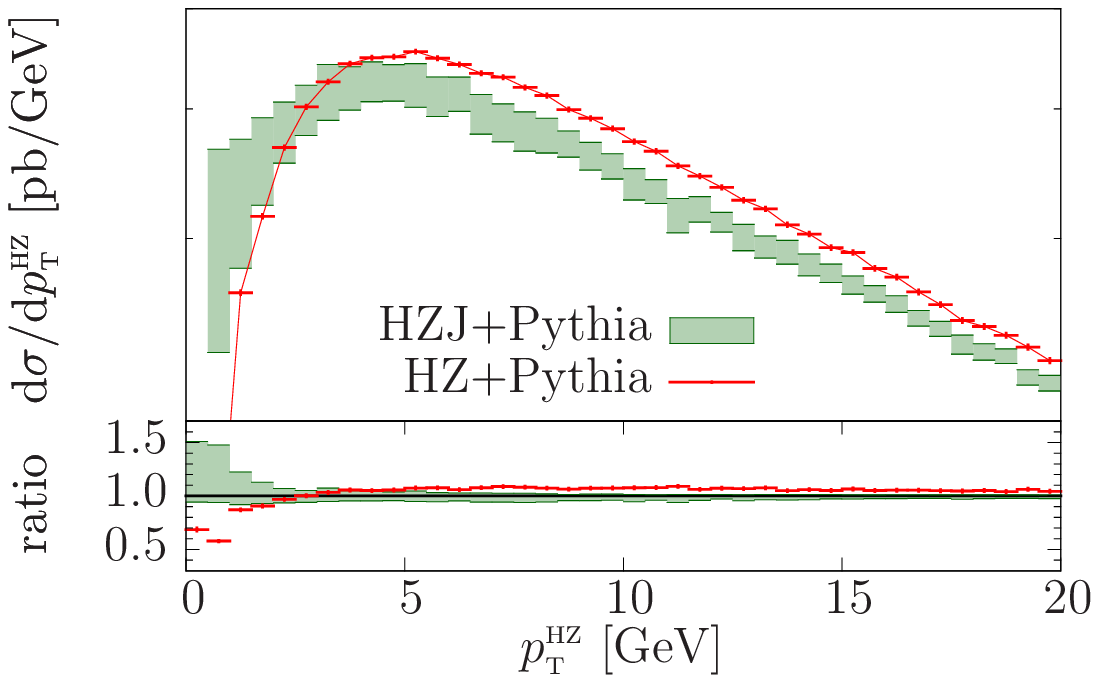}
\end{center}
\caption{Same as fig.~\ref{fig:HZpt} for a different $\pt^{\rm\sss HZ}$
  range.}
\label{fig:HZptzm} 
\end{figure}

\begin{figure}[htb]
\begin{center}
\includegraphics[width=0.49\textwidth]{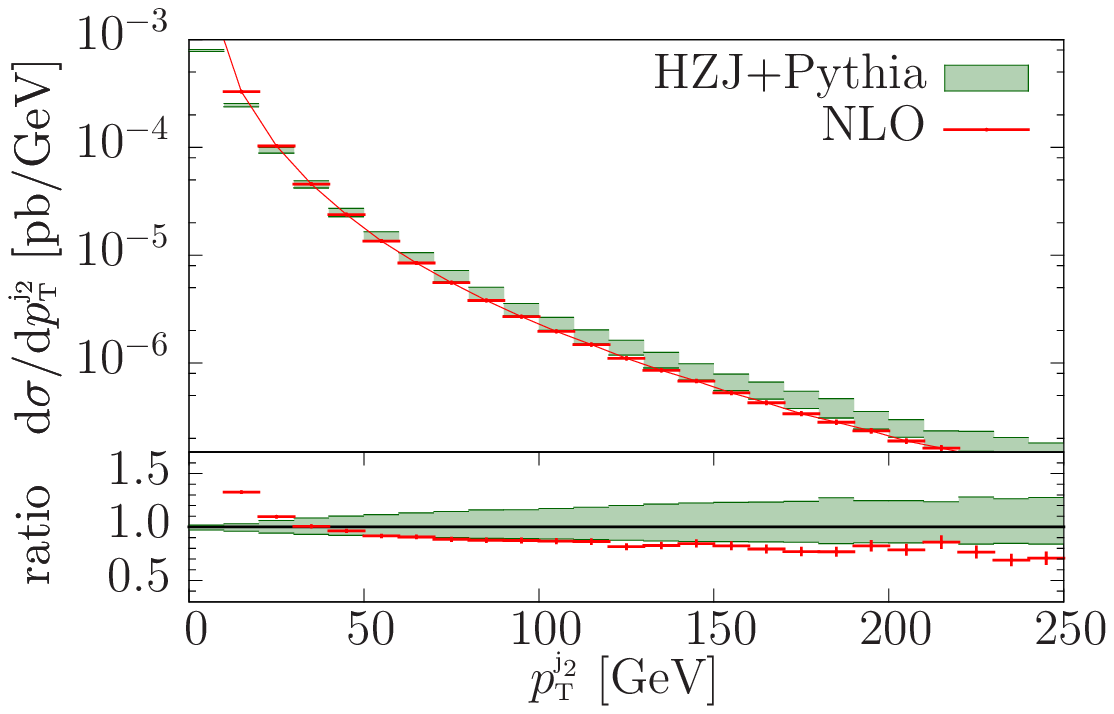}
\includegraphics[width=0.49\textwidth]{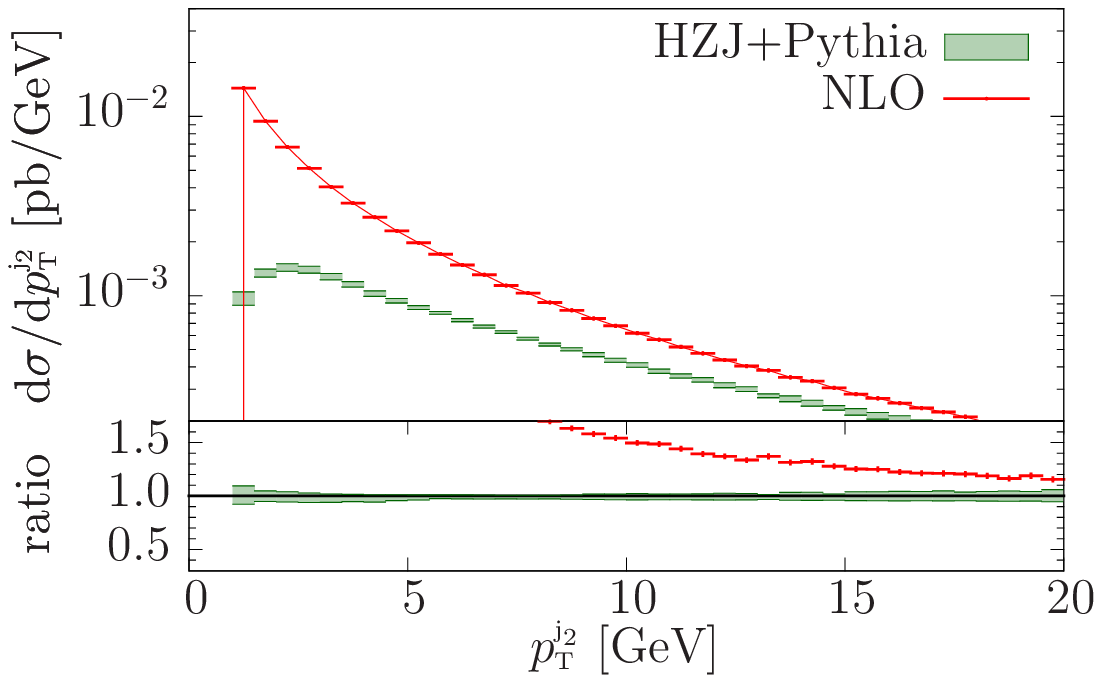}
\end{center}
\caption{Comparison between the \HZJ-\MiNLO+\PYTHIA{} and the NLO \HZJ{}
  result for the transverse momentum of the second hardest jet, at the LHC at
  8~TeV, in two different $\pt$ ranges.  The plots shows the 7-point
  scale-variation band for the \HZJ{} generator.}
\label{fig:HZ_j2pt} 
\end{figure}

Conclusions similar to those for $HW(j)$ can be drawn for $HZ(j)$ associated
production. For this reason, we refrain from commenting
figs.~\ref{fig:HZy}--\ref{fig:HZ_j2pt} that show the same physical quantities
shown previously but for $HZ$ production.

\section{Conclusions}
\label{sec:conclusions}
In this paper we presented new \POWHEGBOX{} generators for $HV$ and $HV$ + 1
jet production, with all spin correlations from the vector boson decay into
leptons correctly included.  The codes for the Born and the real
contributions were computed using the existing interface to \MG{}, while the
code for the virtual amplitude was computed using a new interface to
\GOSAM{}.  This interface allows for the automatic generation of the virtual
code for generic Standard Model processes. With the addition of this new
tool, the generation of all matrix elements in the \POWHEGBOX{} package is
performed automatically, and one only needs to supply the Born phase space in
order to build a \POWHEG{} process.

We have applied the recently proposed \MINLO{} procedure to the
\POWHEG{} $HVj$ production in order to have a generator that is NLO accurate
not only for inclusive distribution in $HVj$ production (as a \POWHEG{}
process is) but also in $HV$ production, i.e.~when the associated jet is not
resolved.  Together with $H/W/Z$ production described in
ref.~\cite{Hamilton:2012rf}, this is a further example of matched calculation
with no matching scale.

We have found very good agreement between the \HVJ{}+\MINLO{} results and the
\HV{} ones, for $HV$ inclusive distributions, while there are clearly
differences in the less inclusive distributions, where the  \HV{} code has at
most leading-order+parton-shower accuracy, while the \HVJ{} one reaches
next-to-leading order accuracy.

We point out that, using our \HVJ{}-\MiNLO{} generator, it is actually
possible to construct an \NNLOPS{} generator, simply by reweighting the
transverse-momentum integral of the cross section to the one computed at the
NNLO level. We postpone a phenomenological study of this method to a future
publication.

\section*{Acknowledgments}
\label{sec:Acknowledgments}

We acknowledge several fruitful discussions with the other members of the
\GOSAM{} collaboration.  The work of G.L. was supported by the Alexander von
Humboldt Foundation, in the framework of the Sofja Kovaleskaja Award Project
``Advanced Mathematical Methods for Particle Physics'', endowed by the German
Federal Ministry of Education and Research. G.L. would like to thank the
University of Milano-Bicocca for support and hospitality during the early
stages of the work, and the CERN PH-TH Department for partial support and
hospitality during later stages of the work.

\appendix

\section{Generation of a new process using \GOSAM{} within the \POWHEGBOX{}}
\label{app:gosam_how_to}

In order to generate a new process in the \POWHEGBOX{} using \GOSAM{}, the
user has to install \QGRAF{}~\cite{Nogueira:1991ex},
\FORM{}~\cite{Kuipers:2012rf}, in addition to the \GOSAM{} package.

The first step in the generation of the code for a new process is to create a
directory under the main \POWHEGBOX{}, and to work from inside this folder,
from where all the following script files have to be executed.  We will refer
to this directory as the \textit{process folder}.  For a complete generation
of a new process the following basic steps are needed:
\begin{itemize}

 \item Generate the tree-level amplitudes and the related code using
   \MG{}~\cite{Campbell:2012am}. After having copied a \MADGRAPH{} input
   card (\texttt{proc\_card.dat}) into the process folder, it is sufficient
   to run from there the \texttt{BuildMad.sh} script contained in the \MGS{}
   folder distributed within the \POWHEGBOX{}. Among the many files
   generated, this will automatically generate an order file for \GOSAM.

 \item Generate the one-loop amplitudes by running the script
   \texttt{BuildGS.sh} contained in the \GSS{} folder distributed within the
   \POWHEGBOX{} with the tag \texttt{virtual}. The script looks for a
   \GOSAM{} input card within the process folder. If no card is found, the
   user is asked if the template one should be used instead. This
   command generates all the needed code for the evaluation of the virtual
   amplitude.

 \item Generate the interface of the virtual code to the \POWHEGBOX. This is
   done by running again the script \texttt{BuildGS.sh} with the tag
   \texttt{interface}.  This replaces the files \texttt{init\_couplings.f}
   and \texttt{virtual.f} with new ones, containing calls to set the values
   of the physical parameters and the initialization of the \GOSAM-generated
   virtual amplitudes.
   
   The parameters are passed by the \POWHEGBOX{} using the function
   \texttt{OLP\_OPTION}, which is not part of the BLHA standards, but is
   described in the \GOSAM{} manual~\cite{Gosammanual}. The new file
   \texttt{virtual.f} instead is constructed using the information contained
   in the contract file, in such a way that the partonic subprocess label
   assigned by \GOSAM{} is not read from the contract file at every run.

 \item Since the evaluation of the virtual amplitude at running time is
   performed with \SAMURAI{} and \GOLEM{}, both distributed in the
   \GOSAMCONT{} package, the last step consists in producing a standalone
   version of everything needed to compute the one-loop amplitudes. This
   can be achieved by executing a last time the script \texttt{BuildGS.sh}
   with the tag \texttt{standalone}. The entire code generated by \GOSAM{},
   together with the code contained in the \GOSAMCONT{} package, is copied in
   the directory \texttt{GoSamlib}, that is then ready to be compiled
   together with the rest of the code. The last three steps can be executed
   all together by running \texttt{BuildGS.sh} with the tag \texttt{allvirt}.
\end{itemize}
If the generation is successful, the only work left to the user is to provide
an appropriate Born phase-space generator in the \texttt{Born\_phsp.f} file.

\section{The \GOSAM{} input card}
\label{app:input_card}
In this section, we describe the main options of the template \GOSAM{} input
card available in the directory \GSS\texttt{/Templates} of the
\POWHEGBOX{} distribution. Further input options for \GOSAM{} can be found in
the online manual~\cite{Gosammanual}. There are two categories of input
options: one related to the characteristics of the physics process and one
more related to the computer code.

\subsection{Physics option}
We list here the main options related to the physics of the processes. For
further options we refer to the \GOSAM{} manual.

\begin{description}

 \item[{\bf model:}] the first thing to choose is the model needed. By
   default, \GOSAM{} offers the choice among the following three models:
   \texttt{sm}, \texttt{smdiag} and \texttt{smehc}, which refer to the
   Standard Model, the Standard Model with diagonal CKM matrix and the
   Standard Model with effective Higgs-gluon-gluon coupling respectively.

% \item[{\bf model.options:}] this tag allows to set some specific model
%   options or some model parameters from the beginning. Since the values of
%   the parameters are nevertheless passed dynamically at the begining of every
%   run from the \POWHEGBOX{} to the code evaluating the virtual amplitudes,
%   this option is not strictly needed. For this reason it is commented out in
%   the template input card.

 \item[{\bf one:}] to simplify the algebraic expressions of the virtual
   amplitudes, a list of parameters can be set algebraically to one using
   this tag. It will not be possible to change the value of the parameter in
   the generated code, since the latter will not contain a variable for the
   corresponding parameter any longer. Due to the normalization conventions
   used between \GOSAM{} and the \POWHEGBOX{}, the virtual amplitudes are
   always returned stripped off from the strong coupling constant
   \texttt{$g_{s}$} and the electromagnetic coupling \texttt{$e$}, which are
   therefore set to one using this tag.

 \item[{\bf zero:}] this tag is similar to the previous one and allows to
   set parameters equal to zero. It is useful to set to zero the desired
   quark and lepton masses as well as the resonance widths.

 \item[{\bf symmetries:}] this tag specifies some further symmetries in the
   calculation of the amplitudes. The information is used when the list of
   helicities is generated. Possible values are:
   \begin{itemize}
   \item \texttt{flavour}: does not allow for flavour changing
     interactions. When this option is set, fermion lines %of PDGs 1-6 
     are assumed not to mix.
   \item \texttt{family}: allows for flavour-changing interactions only
     within the same family. When this option is set, fermion lines %of PDGs
     1-6 are assumed to mix only within families. This means that e.g.~a
     quark line connecting an up with down quark would be considered, while a
     up-bottom one would not.
   \item \texttt{lepton}: means for leptons what ``flavour'' means for
     quarks.
   \item \texttt{generation}: means for leptons what ``family'' means for
     quarks.
   \end{itemize}
   Furthermore it is possible to fix the helicity of particles. This can be
   done using the command \texttt{\%<n>=<h>}, where $<n>$ stands for a PDG
   number and $<h>$ for an helicity. For example \texttt{\%23=+-} specifies
   the helicity of all $Z$-bosons to be ``+'' and ``-'' only (no ``0''
   polarisation).

 \item[{\bf qgraf.options:}] this is a list of options to be passed to
   \QGRAF{}. For the complete set of possible options we refer to the
   \QGRAF{} manual~\cite{Nogueira:1991ex}. Customary options which are used are
   \texttt{onshell}, \texttt{notadpole}, \texttt{nosnail}.

 %% \item[{\bf qgraf.verbatim:}] this tag allows to send verbatim lines to the
 %%   file qgraf.dat. Some examples of verbatim lines are given in the default
 %%   \gosamrc{} file. Further possibilities can be found in the \QGRAF{}
 %%   manual~\cite{Nogueira:1991ex}.

 \item[{\bf filter.module, filter.lo, filter.nlo:}] these tags can impose
   user-defined filters to the LO and NLO diagrams, passed to \GOSAM{} via a
   python function. Some ready-to-use filter functions are provided by
   default by \GOSAM{}. A complete list can be found in the \GOSAM{} online
   manual. Nevertheless, further filters can be constructed combining
   existing ones. Ideally, these filters are defined in a separate file,
   which we call \texttt{filter.py}. The tag \texttt{filter.module} can be
   used to set the PATH to the file containing the filter.

As an example we report here the filters used for the generation of the codes
of the processes presented in this paper, where we want to neglect diagrams
in which the Higgs boson couples directly to the massless
fermions. Therefore, in a file called \texttt{filter.py}, we define the
following filter, selecting diagrams which do not have this vertex:
\\~\\
\texttt{
def no\_hff(d):\\
   """\\
   No Higgs attached to massless fermions.\\
   """\\
   return d.vertices([H], [U,D,S,C,B], [Ubar,Dbar,Sbar,Cbar,Bbar]) == 0\\~\\
   }
In the \gosamrc{} card we can then add the following lines:\\
\texttt{
filter out filter.module=filter.py\\
filter.lo= no\_hff\\
filter.nlo= no\_hff
}

 \item[{\bf extensions:}] this tag can be used to list a set of options to be
   applied in the generation of the one-loop code. Among them, the name of
   the code that will be used in the computation of the diagrams at running
   time (usually \SAMURAI{} and \GOLEM{}), if numerical polarization vectors
   for external massless vector bosons should be used, if the code should be
   generated with the option that the Monte Carlo programs controls the
   numbering of the files containing unstable points\ldots The list of
   possible extensions useful in conjunction with the \POWHEGBOX{} is:
   \begin{itemize}
   \item \texttt{samurai}: use \SAMURAI{} for the reduction
   \item \texttt{golem95}: use \GOLEM for the reduction
   \item \texttt{numpolvec}: evaluate polarization vectors numerically
   \item \texttt{derive}: tensorial reconstruction using derivatives
   \item \texttt{formopt}: (with \FORM{} $\ge$ 4.0) diagram optimization
     using \FORM{} (works only with \texttt{abbrev.level=diagram})
% and \texttt{r2=implicit/explicit})
   \item \texttt{olp\_badpts}: (OLP interface only): allows to control the
     numbering of the files containing bad points from the MC program. This
     should always be used when generating a code for the \POWHEGBOX{}.
   \end{itemize}

 \item[{\bf PSP\_check:}] this tag switches the detection of unstable points
   on and off and can take the values \texttt{true} or \texttt{false}.

 \item[{\bf PSP\_verbosity:}] the verbosity of the {\bf PSP\_check} can be set
   here. The possible values are:\\ 
   \texttt{verbosity = 0} : no output,\\ 
   \texttt{verbosity = 1} : bad points are written in a file stored in the
   folder \texttt{BadPoints}, which is automatically created in the process
   folder.\\   
   \texttt{verbosity = 2} : output whenever the rescue system is used with
   comments about the success of the rescue. 

 \item[{\bf PSP\_chk\_threshold1, PSP\_chk\_threshold2:}] tags to set the
   threshold used to declare if a point is unstable or not. These thresholds
   are integers, indicating the number of digits of precision which are
   required. The first threshold acts on the result given by \SAMURAI{}. If a
   phase-space point does not fulfill the required precision, it is
   recomputed using \GOLEM{}. The second threshold acts on the result from
   \GOLEM{}. If also the result of \GOLEM{} does not fulfill the required
   accuracy, the phase-space point and some further information are written
   in the \texttt{BadPoints} folder, provided the verbosity flag is set.

 \item[{\bf PSP\_chk\_kfactor:}] a further threshold on the $K$-factor of the
   virtual amplitude can be set using this tag. If the value is set negative,
   this tag has no effect and all $K$-factor values are accepted.

 \item[{\bf diagsum:}] to increase the speed of the evaluation of the virtual
   matrix elements, \GOSAM{} can add diagrams which share identical
   loop-propagators and differ only in the external tree-structure attached
   to the loop, before the algebraic reduction takes place. This option can
   be switched on and off by setting this tag to \texttt{true} or
   \texttt{false}.

 \item[{\bf abbrev.level:}] for an optimal computational speed, during the
   preparation of the one-loop code, \GOSAM{} groups identical algebraic
   structures into abbreviations. This tag allows to set at which level these
   abbreviations should be defined. Possible values are \texttt{helicity},
   \texttt{group} and \texttt{diagram}. The \texttt{helicity} level is most
   indicated for easy processes with a small number of diagrams. If the
   extension \texttt{formopt} is used, the \texttt{abbrev.level}
   \textit{must} be set to \texttt{diagram}.
\end{description}

\subsection{Computer option}
Contrary to a standalone use of \GOSAM{}, within the \POWHEGBOX{} framework a
separate installation of the \texttt{gosam-contrib} package is not necessary,
since this is contained in the \POWHEGBOX{} distribution. The following tags
allow to set some options for the external programs:
\begin{description}
 \item[{\bf qgraf.bin:}] the location of the \QGRAF{} executable.
 \item[{\bf form.bin:}] the location of the \FORM{} executable. Usually the
   user can choose between the standard version (\texttt{form}) and the
   multi-threads version called \texttt{tform}.
 \item[{\bf form.threads:}] when using the multi-thread version of
   \FORM, the number of threads to be used can be set here.
 \item[{\bf form.tempdir:}] the path to the folder where \FORM{} saves
   temporary files can be set using this tag.
\end{description}

\providecommand{\href}[2]{#2}\begingroup\raggedright\endgroup

\end{document}